\documentstyle[11pt,aaspp4]{article}
\def\lineunits{ergs\ s$^{-1}$\,cm$^{-2}$}
\def\contunits{ergs\ s$^{-1}$\,cm$^{-2}$\,\AA$^{-1}$}
\def\Fvar{\ifmmode F_{\rm var} \else $F_{\rm var}$\fi}
\def\Rmax{\ifmmode R_{\rm max} \else $R_{\rm max}$\fi}
\def\tcent{\ifmmode \tau_{\rm cent} \else $\tau_{\rm cent}$\fi}

\def\deg{\ifmmode ^{\rm o} \else $^{\rm o}$\fi}

\def\arcsecpoint{\ifmmode ''\!. \else $''\!.$\fi}

\def\kms{\ifmmode {\rm km\ s}^{-1} \else km s$^{-1}$\fi}
\def\Msun{\ifmmode M_{\odot} \else $M_{\odot}$\fi}
\def\Lsun{\ifmmode L_{\odot} \else $L_{\odot}$\fi}
\def\Rs{\ifmmode R_{\rm S} \else $R_{\rm S}$\fi}
\def\qo{\ifmmode q_{\rm o} \else $q_{\rm o}$\fi}
\def\Ho{\ifmmode H_{\rm o} \else $H_{\rm o}$\fi}
\def\ho{\ifmmode h_{\rm o} \else $h_{\rm o}$\fi}
\def\ltsim{\raisebox{-.5ex}{$\;\stackrel{<}{\sim}\;$}}
\def\gtsim{\raisebox{-.5ex}{$\;\stackrel{>}{\sim}\;$}}
\def\vFWHM{\ifmmode V_{\mbox{\tiny FWHM}} \else
            $V_{\mbox{\tiny FWHM}}$\fi}
\def\CCF{\ifmmode F_{\it CCF} \else $F_{\it CCF}$\fi}
\def\ACF{\ifmmode F_{\it ACF} \else $F_{\it ACF}$\fi}
\def\Halpha{\ifmmode {\rm H}\alpha \else H$\alpha$\fi}
\def\Hbeta{\ifmmode {\rm H}\beta \else H$\beta$\fi}
\def\4686{\ifmmode {\mbox{\rm He\,}{\sc ii}}\,\lambda4686
	\else He\,{\sc ii}\,$\lambda4686$\fi}
\def\Hgamma{\ifmmode {\rm H}\gamma \else H$\gamma$\fi}
\def\Hdelta{\ifmmode {\rm H}\delta \else H$\delta$\fi}
\def\Lya{\ifmmode {\rm Ly}\alpha \else Ly$\alpha$\fi}
\def\Lyb{\ifmmode {\rm Ly}\beta \else Ly$\beta$\fi}

\def\heii{He\,{\sc ii}}

\def\ciii{\ifmmode {\rm C}\,{\sc iii} \else C\,{\sc iii}\fi}
\def\civ{\ifmmode {\rm C}\,{\sc iv} \else C\,{\sc iv}\fi}

\def\oiii{O\,{\sc iii}}
\def\o5007{[O\,{\sc iii}]\,$\lambda5007$}

\def\feii{Fe\,{\sc ii}}

\lefthead{Peterson et al.}
\righthead{Variability in NGC 4051}

\begin{document}
\title{X-Ray and Optical Variability in NGC~4051
and the Nature of Narrow-Line Seyfert 1 Galaxies }

\author{
B.M.~Peterson,\altaffilmark{1}
I.M.~McHardy,\altaffilmark{2}
B.J.~Wilkes,\altaffilmark{3}
P.~Berlind,\altaffilmark{3}
R.~Bertram,\altaffilmark{1}$^{,}$\altaffilmark{4}
M.~Calkins,\altaffilmark{3}
S.J.~Collier,\altaffilmark{1}
J.P.~Huchra,\altaffilmark{3}
S.~Mathur,\altaffilmark{1}
I.~Papadakis,\altaffilmark{5}
J.~Peters,\altaffilmark{3}$^{,}$\altaffilmark{6}
R.W.~Pogge,\altaffilmark{1}
P.~Romano,\altaffilmark{1}
S.~Tokarz,\altaffilmark{3}
P.~Uttley,\altaffilmark{2}
M.~Vestergaard,\altaffilmark{1}
and R.M.~Wagner\altaffilmark{1}$^{,}$\altaffilmark{4}
}

\altaffiltext{1}
        {Department of Astronomy, The Ohio State University,
                140 West 18th Avenue, Columbus, OH  43210.
        \\ Email: peterson, stefan, smita, pogge, promano, 
	vester@astronomy.ohio-state.edu}
\altaffiltext{2}
	{Department of Physics and Astronomy, University of Southampton,
	Southampton SO17 1BJ, United Kingdom.
	\\ Email: imh, pu@astro.soton.ac.uk}
\altaffiltext{3}
        {Harvard-Smithsonian Center for Astrophysics,
        60 Garden Street, Cambridge, MA 02138.
        \\ Email: belinda, pberlind, mcalkins,
	huchra, stokarz@cfa.harvard.edu}
\altaffiltext{4}
        {Mailing address: Steward Observatory, University of Arizona,
	Tucson, AZ 85721. \\
	Email: rayb, rmw@as.arizona.edu}
\altaffiltext{5}
        {Department of Physics, University of Crete,
	P.O.\ Box 2208, 71 003 Heraklion, Crete, Greece.\\
	Email: jhep@physics.uoc.gr}
\altaffiltext{6}
	{Deceased.}

\begin{abstract}
We report on the results of a three-year program of coordinated
X-ray and optical monitoring of the narrow-line Seyfert 1 galaxy
NGC~4051. The rapid continuum variations observed in the X-ray
spectra are not detected in the optical, although the 
{\it time-averaged} X-ray and optical continuum fluxes are well-correlated.
Variations in the flux of the broad \Hbeta\ line
are found to lag behind the optical continuum variations by
6 days (with an uncertainty of 2--3 days), and combining
this with the line width yields a virial mass estimate of
$\sim1.1 \times 10^6$\,\Msun, at the very low end of the distribution
of AGN masses measured by line reverberation. Strong variability of
\4686\ is also detected, and the response time measured is similar to
that of \Hbeta, but with a much larger uncertainty. The
\4686\ line is almost five times broader than \Hbeta, and it is
strongly blueward asymmetric, as are the high-ionization UV lines
recorded in archival spectra of NGC~4051. The data are consistent
with the Balmer lines arising in a low to moderate inclination 
disk-like configuration, and the high-ionization lines arising in
an outflowing wind, of which we observe preferentially the near side.
Previous observations of the narrow-line region morphology 
of this source suggest that the system is
inclined by $\sim50$\deg, and if this is applicable to the
broad \Hbeta-emitting region, a central mass of 
$\sim1.4 \times 10^6$\,\Msun\ can be inferred.
During the third year of monitoring,
both the X-ray continuum and the \4686\ line went into extremely
low states, although the optical continuum and the \Hbeta\ broad line
were both still present and variable. We suggest that the inner
part of the accretion disk may have gone into an advection-dominated
state, yielding little radiation from the hotter inner disk.
\end{abstract}

 
\setcounter{footnote}{0}

\section{Introduction}
The broad components of permitted emission lines in the spectra
of active galactic nuclei (AGNs) typically have velocity
widths of a few to several thousands of kilometers per second.
The defining characteristic of the subclass of AGNs known as 
narrow-line Seyfert 1 galaxies (NLS1s) 
is that the broad components of their emission lines are
much narrower ($\vFWHM \ltsim 2000$\,\kms) than is typical
for broad-line objects (Osterbrock \& Pogge 1985). 
NLS1s are extreme AGNs in other respects as well ---
their UV--optical properties correlate well with the
Boroson \& Green (1992) primary spectral eigenvector identified in
principal component analysis. In other words, NLS1 classification
correlates well with strong optical \feii\ emission and weak
[\oiii]\,$\lambda\lambda4959$, 5007 emission (Boller, Brandt \& Fink 1996).
While the possible importance of such extreme AGNs has been
recognized for two decades (Davidson \& Kinman 1978),
interest in NLS1s has increased recently as their unusual X-ray
properties have come to light: they have unusually steep 
soft and hard X-ray
spectra (Puchnarewicz et al. 1992; Boller, Brandt, \& Fink 1996;
Brandt, Mathur, \& Elvis 1997) and undergo rapid
non-linear variability (Boller et al.\ 1997). While rapid,
large amplitude variability in the UV--optical has not been
reported, existing data do not address well
the relationship between the X-ray and long-wavelength
variations. A good compilation
of the observed properties of NLS1s is given by
Taniguchi, Murayama, \& Nagao (1999).

Possible explanations for the narrowness of the permitted lines
include the following:
\begin{enumerate}
\item NLS1s have more distant-than-normal BLRs (Mason, Puchnarewicz, \& Jones
1996; Wandel \& Boller 1998). The line widths are attributed
to virial motion ($M \approx V^2r/G$)
in the vicinity of a supermassive black hole,
but the orbital velocities are smaller than in more typical AGNs
because of greater distance to the central source.
\item NLS1s are low-inclination (i.e., nearly face-on)
systems (Osterbrock \& Pogge 1985).
In this model, the line widths are again due to orbital motion around
the central black hole, and the bulk of the broad-line region (BLR)
gas orbits in a common plane that is almost perpendicular to
the line of sight, leading to relatively small Doppler
widths.
\item NLS1s have relatively low black-hole masses, but
high accretion rates. Again, the basic assumption is that
the BLR motions are virial, but the central source has
a lower mass. The luminosity can be kept relatively high
by supposing that the accretion rate (relative to the
Eddington rate) is correspondingly high in these sources.
\end{enumerate}
The third of these explanations forms the current paradigm
for NLS1s, as it explains not only the narrow emission lines,
but possibly also the steepness of the soft X-ray spectrum:
the temperature structure of an optically thick, geometrically
thin accretion disk is given by
\begin{equation}
T(r) = 6.3 \times 10^5 \left(\dot{M}/\dot{M}_{\rm Edd}\right)^{1/4}
M_8^{-1/4} \left(r/R_{\rm S} \right)^{-3/4}\ {\rm K,}
\end{equation}
where $M_8$ is the black hole mass in units of $10^8\,M_{\odot}$
and $R_{\rm S}$ is the Schwarzschild radius
(Shakura \& Sunyaev 1973).
The temperature of the inner regions of the disk scales
like $M^{-1/4}$, so the strength of the soft X-rays in NLS1s
might plausibly be ascribed to a low central mass and
a high accretion-rate flow.

One way to distinguish among various explanations of the NLS1
phenomenon is to measure the size of the BLR, which can be
done via reverberation-mapping techniques (Blandford \& McKee 1982;
Peterson 1993; Netzer \& Peterson 1997). It is usually assumed
that the variability of an emission-line $L(t)$ to a variable
continuum $C(t)$ can be linearized to the form
\begin{equation}
\label{eq:TF}
\delta L(t) = \int \Psi(\tau)\ \delta C(t-\tau)\ d\tau,
\end{equation}
where $\Psi(\tau)$ is the ``transfer function,'' which depends on
the geometry and reprocessing physics of the BLR.
A representative time scale for response of a line can be found
by a simple cross-correlation of the  continuum and emission-line light
curves. By convolving eq.\ (\ref{eq:TF}) with $C(t)$, one obtains
the cross-correlation function
\begin{equation}
\CCF(\tau) = \int \Psi(\tau') \ACF(\tau-\tau')\ d\tau',
\end{equation}
where $\ACF(\tau)$ is the continuum autocorrelation function
(Penston 1991; Peterson 1993). The cross correlation lag $\tau$
can be taken to be the light-travel time across the BLR, so
the BLR size is given by $r=c\tau$.
By combining this with the emission-line width $\vFWHM$,
the mass of the central source can be inferred to be
\begin{equation}
M = \frac{f \vFWHM^2 c\tau}{G},
\end{equation}
where $f$ is a factor of order unity that depends on the still
unknown geometry and kinematics of the BLR. There is an
implicit assumption that the gravitational force of the
central object dominates the kinematics of the BLR; this is
formally unproven, but at least in the case of the well-studied
Seyfert 1 galaxy NGC~5548, the reverberation-mapping data are
consistent with the required $\vFWHM \propto r^{-1/2}$ relationship
(Peterson \& Wandel 1999). Virial mass estimates based on
reverberation-mapping data are now available
for nearly 40 AGNs (Wandel, Peterson, \& Malkan 1999;
Kaspi et al.\ 2000).

Beginning in early 1996, we undertook a program of 
contemporaneous X-ray and optical spectroscopic
monitoring of the galaxy NGC~4051, the only NLS1
galaxy in Seyfert's (1943) original list of high surface-brightness
galaxies with strong emission lines. The X-ray variability
characteristics of NGC~4051 are typical of the NLS1 class
(Lawrence et al.\ 1987; McHardy et al.\ 1995).

X-ray observations were made 
with the {\em Rossi X-Ray Timing Explorer (RXTE)},
and optical spectra were obtained with ground-based
telescopes, as described below. The purpose
of this program has been twofold:
\begin{enumerate}
\item To determine the nature of the relationship between the
X-ray and UV--optical continuum variations. This is
a particularly interesting question in the case of NGC~4051
since the X-ray flux dropped to an extremely low level
in 1998 May (Uttley et al.\ 1999), towards the end of this campaign.
\item To determine the BLR size and virial mass via
reverberation techniques.
\end{enumerate}
In this contribution, we present the results of this
program, and discuss their implications for the nature
of the NLS1 phenomenon.

\section{Observations and Data Analysis}

\subsection{X-Ray Observations}

The X-ray observations were made with the large area (0.7 m$^{2}$)
Proportional Counter Array (PCA) on {\it RXTE}
(Bradt, Rothschild \& Swank 1993). The observations shown
here, which are part of a continuing monitoring program, cover the
period from 1996 March to 1998 June. The observations were scheduled
initially to cover the largest range of variability time scales with
the smallest number of observations in order to determine the X-ray
power spectrum efficiently. The program therefore consisted of
observations approximately every 7 to 10 days throughout the first
year, followed by observations approximately every 2 weeks
thereafter. In addition, during the first year, there were two more
intensive monitoring  periods: a two-week period of twice-daily
observations and a four-week period of of daily observations.
Each observation lasted for typically $1$\,ksec. In 
1996 December, there was also a period of 3 days during which NGC~4051 was
observed for a total of 70\,ksec.

The PCA consists of 5 proportional counter units (PCUs) but typically
only 3 PCUs (numbers 0, 1 and 2) were operational and all count rates
refer to the total counts from those 3 PCUs.  Where other than 3 PCUs
were in operation, the count rate has been normalized to  3
PCUs.  The PCUs have 3 layers. Here we only include data from the
upper layer as this layer provides the highest 
signal-to-noise ratio for photons in the
energy range 2--20 keV where the flux from NGC~4051 is strongest.  We
used standard ``good time interval'' (GTI) criteria to select data with
the lowest background.  We reject data obtained when the Earth
elevation angle was less than 10$^{\circ}$, when the pointing offset
from the source was $>0.02^{\circ}$, or during passage through the
South Atlantic Anomaly, or up to 5 minutes afterwards.  The PCA is a
non-imaging device with a field of view of FWHM $\sim1^\circ$ and so
the background which we subtract must be calculated from a model.
Here we use the FTOOL routine PCABACKEST V2.0c, with the new ``L7''
model, to calculate the background.

The resultant 2--10 keV lightcurve is shown in the top panel of
Fig.\ 1. Further details of the X-ray light curves and
variability are given by Papadakis et al.\ (2000).
For NGC~4051, 10 counts s$^{-1}$ (2--10 keV),  
from 3 PCUs, corresponds to a flux of 
$4 \times 10^{-11}$ ergs cm$^{-2}$ s$^{-1}$.

\subsection{Optical Spectroscopy}

Optical spectroscopic observations were obtained 
between UT 1996 January 12 (Julian Date = JD2450095) and
1998 July 28 (JD2451022), covering three 
separate observing seasons. Observations were made with
the Ohio State CCD Spectrograph on the 1.8-m Perkins Telescope
of the Ohio State and Ohio Wesleyan Universities at the
Lowell Observatory (data set A) and with the FAST spectrograph
(Fabricant et al.\ 1998)
on the 1.5-m Tillinghast Reflector of the Center for
Astrophysics on Mt.\ Hopkins (data set B). A log of
these observations is presented in Table 1. Column (1) gives the
UT date of each observation and the Julian Date is given in column (2).
The origin of the data (set A or B) appears in column (3). 
The projected size of the spectrograph entrance 
aperture was $5\arcsecpoint0\times7\arcsecpoint5$ 
(i.e., a slit width of $5\arcsecpoint0$ and 
a cross-dispersion extraction window of 
$7\arcsecpoint5$) for all set A spectra
and $3\arcsecpoint0\times4\arcsecpoint6$ for all set B spectra.
In each case, the slit was oriented in the east-west
direction (i.e., the slit position angle was always 90\deg).
The nominal spectral resolution was 9\,\AA\ for all set A spectra and
5\,\AA\ for all set B spectra. The wavelength coverage of each 
spectrum is given in column (4), and the file name
is given in column (5). All of these spectra are publicly
available on the International AGN Watch 
site on the World-Wide Web\footnote{The light curves 
and spectra are available at URL 
{\sf http://www.astronomy.ohio-state.edu/$\sim$agnwatch/}.
All publicly available International AGN Watch data can be accessed
at this site, which also includes complete references to
published AGN Watch papers.}.

The spectroscopic images were processed in standard
fashion for CCD frames, including bias subtraction, 
dark-count correction when necessary,
flat-field correction, wavelength calibration,
and flux calibration based on standard-star observations.
Since even under photometric conditions, AGN spectrophotometry is rarely more
accurate than $\sim$10\%, the usual technique of flux calibration by
comparison with standard stars is far too poor for
AGN variability studies. We therefore base our flux calibration
on a scale defined by the observed flux in the prominent
narrow [O\,{\sc iii}]\,$\lambda\lambda4959,$ 5007 doublet.
These lines originate in a low-density region that is more spatially
extended than the BLR or the continuum source, which for
practical purposes can be regarded as point sources.
The larger light-travel time and long recombination time
ensure that any narrow-line variations will occur only over
much longer time scales than of interest in this experiment.
We therefore assume that the [O\,{\sc iii}] lines are constant
in flux, and use these to scale each spectrum.  All spectra
are scaled to a constant flux of
$F(\mbox{\o5007}) = (3.91 \pm 0.12) \times 10^{-13}$\,\lineunits,
which is based on the mean of ten spectra from data set A that were
obtained under photometric observing conditions during the 1996
observing season. The \o5007\ flux measured from photometric
spectra from subsequent years supports our assumption that this
value can be assumed to be constant over the time scales of interest.
All of the 
spectra are adjusted in flux to have this value of $F(\mbox{\o5007})$
by employing the spectral scaling software described
by van Groningen \& Wanders (1992). The process is as follows:
spectra are adjusted
in flux by a multiplicative constant that is determined by comparing each
spectrum to a ``reference spectrum'' that has been formed
by averaging all the highest-quality (i.e., typically
signal-to-noise ratios $S/N \gtsim 30$) spectra and scaling this
mean spectrum to the adopted \o5007\ flux. All individual 
spectra are scaled relative to the reference spectrum in a least-squares
fashion that minimizes the [\oiii] residuals in the difference
spectrum produced by subtracting the reference spectrum from
each individual spectrum. This  
program also corrects for small zero-point
wavelength-calibration errors between the
individual spectra, and takes resolution differences into account. 

At this point, measurements of each of the spectra are made.
The continuum flux at $\sim5100$\,\AA\ (in the rest frame of
NGC 4051, $z = 0.002418$, based on 21-cm emission 
[deVaucouleurs et al.\ 1991]) 
is determined by averaging the flux in 
the 5090--5120\,\AA\ bandpass (in the observed frame).
The \Hbeta\ emission-line flux is measured by assuming a linear
underlying continuum between $\sim4772$\,\AA\ and $\sim5105$\,\AA,
and integrating the flux above this continuum between 4820\,\AA\ and
4910\,\AA\ (all wavelengths in the observed frame). The long-wavelength
cutoff of this integration band is chosen to avoid the
\feii\ contamination underneath [\oiii]\,$\lambda4959$. We also note
that no attempt has been made to correct for contamination of
the line measurement by the {\em narrow-line}\ component of 
\Hbeta, which is of course expected to be constant. 

We have also measured the flux in the \4686\ line. 
Only set B spectra are suitable for this measurement,
since set A spectra do not extend shortward far enough
to provide a suitable short-wavelength continuum point.
The \heii\ flux was measured by adopting a linear
underlying continuum between $\sim4447$\,\AA\
and $\sim4775$\,\AA, and integrating the flux above this continuum
between 4613\,\AA\ and 4772\,\AA.

We then compare the independent light curves from the two
sets of data to identify small systematic flux differences
between the sets, as we have done in many
previous experiments (see Peterson et al.\ 1999 and references
therein). We attribute these small relative flux offsets
to aperture effects, although the
procedure we use also corrects for other unidentified systematic
differences between data sets.  We
define a point-source correction factor $\varphi$ by the equation
\begin{equation}
\label{eq:defphi}
F(\Hbeta)_{\rm true} = \varphi F(\Hbeta)_{\rm observed}.
\end{equation}
This factor accounts for the fact that different apertures
result in different amounts of light loss for the 
point-spread function (which describes the surface-brightness
distribution of both the broad lines and the AGN continuum
source) and the partially extended narrow-line region (NLR).
After correcting for aperture effects on the point-spread function
to narrow-line ratio, another correction needs to be applied to
adjust for the different amounts of starlight admitted by
different apertures. An extended source correction $G$ 
is thus defined as
\begin{equation}
\label{eq:defG}
F_{\lambda}(5100\,{\textstyle {\rm \AA}})_{\rm true} = \varphi 
F_{\lambda}(5100\,{\textstyle {\rm \AA}})_{\rm observed} - G.
\end{equation}
The value of $G$ is essentially the nominal difference in the 
contaminating host-galaxy flux between the two spectrograph
entrance apertures employed.

This intercalibration procedure is accomplished by
comparing pairs of nearly simultaneous observations from 
the two data sets to determine $\varphi$ and $G$.
In practice, the interval which we define
as ``nearly simultaneous'' is two days or less,
which means that in principle any real variability that occurs on time scales
this short tends to be somewhat suppressed by the
process that allows us to merge the two data sets.
In this case, the adjustment has very little impact on the
final results because nearly all of the  pairs of data
separated by two days or less are from data set B and thus
have not been adjusted relative to one another.
We find that the best-fit constants
for set B relative to set A are $\varphi = 0.982 \pm 0.048$ 
and $G = (-1.243 \pm 0.729) \times 10^{-15}$\,\contunits.
Both the \Hbeta\ and \4686\ fluxes are adjusted as in
eq.\ (\ref{eq:defphi}); the fact that the factor $\varphi$ is so
close to unity indicates that most of the NLR in
this galaxy arises within a few arcseconds of the nucleus.

The final continuum $F_{\lambda}$(5100\,\AA) and \Hbeta\ 
emission-line fluxes are given in Table 2. Simultaneous
(to within 0.1 day) measurements were averaged, weighted by
the reciprocal of their variances.

\section{Analysis}

\subsection{Continuum Variability}

The light curves listed in Table 2 are plotted in Fig.\ 1,
along with contemporaneous 2--10\,keV X-ray light curves.
The data shown here span three observing
seasons, beginning in 1996 January and ending in 
1998 July.  A summary of the general variability characteristics
is given in Table 3. For the complete data base 
and for individual subsets of the data as given in column (1),
columns (2) -- (4) give respectively the number of individual observations
and the average and median time intervals between them.
The mean flux is given in column (5), and columns (6) and (7)
give two widely used measures of the variability,
\Fvar, the root-mean-square (rms) fractional variability 
corrected for measurement error, as defined by 
Rodr\'{\i}guez-Pascual et al.\ (1997), and \Rmax,
the ratio of maximum to minimum flux, respectively.
Both \Fvar\ and \Rmax\  parameters
are affected by contamination of the measured quantities by
constant-flux components; the optical continuum values are
somewhat diluted by the constant contribution of 
the underlying host galaxy, and the emission-line 
values are affected by both narrow-line contributions and
probably slowly varying \feii\ emission as well. In any
case, these will have only a modest effect on \Fvar\ and
\Rmax\ and inspection of Table 3 shows clearly that
the large-amplitude, rapid variations that characterize
X-rays in NLS1s are much less pronounced in the optical
spectrum.

While there is a clear lack of correlated short time-scale
behavior of the X-ray and optical continua, the light curves
in Fig.\ 1 suggest that a correlation on longer time scales
is possible. To test this quantitatively, we have 
suppressed the rapid variations by smoothing both the optical continuum
and X-ray light curves with a rectangular function
of width 30 days, as shown in Fig.\ 3,
similar to what was done by Maoz, Edelson, \& Nandra (2000)
in a comparison of X-ray and optical variability in
the Seyfert 1 galaxy NGC~3516.  Cross-correlation of the
overlapping parts of these 
light curves using the methodology described in the next section
yields a lag of the optical variations relative to the
X-ray of $\tau= 6^{+62}_{-112}$ days with a correlation
coefficient $r_{\rm max} = 0.74$, i.e., the mean X-ray and optical
fluxes are indeed  correlated once the high-frequency variability
is suppressed. The lag between variations in the two wavebands 
is highly uncertain, but consistent with zero or any small
time lag expected in the continuum emitting region.

\subsection{Emission-Line Variability}

Comparison of the optical continuum and emission-line light
curves shows that the variations in each are quite similar,
indicating that the time delay between them is small. The time delay between 
continuum and emission-line variations can be quantified by
cross-correlation of the light curves. During the first year,
there is a period from JD2450183 to JD2450262 in which the
light curves are well-sampled and the character of the variations
permits an accurate cross-correlation measurement. In Table 3,
we refer to this subset of data as ``subset 1'', and we 
plot this part of the light curves in an expanded form in
Fig.\ 2. We cross-correlate the data shown here by using 
the interpolation cross-correlation function (ICCF) method
of Gaskell \& Sparke (1986) and Gaskell \& Peterson (1987) and the
discrete correlation function (DCF) method of Edelson \& Krolik (1988),
where in both cases, we employ the specific  
implementation described by White \& Peterson (1994). 

The results of the cross-correlation analysis are summarized in Table 4
and the cross-correlation functions are shown in Fig.\ 4.
For both \Hbeta\ and \4686, Table 4 gives the ICCF
centroid \tcent, and the location $\tau_{\rm peak}$
of the maximum value of the correlation coefficient
$r_{\rm max}$. The centroid \tcent\
is computed using all points near $\tau_{\rm peak}$ with
values greater than 0.8$r_{\rm max}$. 
The uncertainties quoted for \tcent\ and 
$\tau_{\rm peak}$ are based on the model-independent
Monte-Carlo method described by Peterson et al.\ (1998).

By combining this lag with the Doppler width of the emission
line, we can estimate a virial mass, as in eq.\ (4);
for consistency with Wandel et al.\ (1999) and Kaspi et al.\ (2000),
we use $f=3/\sqrt{2}$ in eq.\ (4).
Since the broad emission-line features are comprised of
a number of different components (or contaminants), it
is desirable to measure the Doppler width of only
the {\em variable} part of the emission line.
In order to isolate the 
variable part of the emission line and exclude constant
components (such as contamination from the NLR),
we measure the relevant line widths in the {\em rms}
spectrum constructed from all the set B spectra in subset 1. 
The mean spectrum is constructed by averaging all $N (= 18)$ 
set B spectra in subset 1, i.e.
\begin{equation}
\bar{F}(\lambda) = \frac{1}{N}
\sum^{N}_{i=1} F_{i}(\lambda),
\end{equation}
where $F_{i}(\lambda)$ is the flux density (in \contunits) at
wavelength $\lambda$ in the $i$th spectrum. The rms spectrum
is similarly constructed as
\begin{equation}
\sigma(\lambda) = \left[ \left( \frac{1}{N-1} \right)
\sum^{N}_{i=1} 
\left( F_{i}(\lambda) - \bar{F}(\lambda)  \right)^{2}
\right]^{1/2}.
\end{equation}
The mean and rms spectra are shown as the top two panels in
Fig.\ 5. 
The widths of the \Hbeta\ and \4686\ emission lines
(full-width at half maximum, \vFWHM) 
are given in Table 4, as Doppler widths in the rest frame
of NGC~4051. A virial mass is then computed as described
by Wandel, Peterson, \& Malkan (1999). On the basis of 
the \Hbeta\ variations, a mass of $1.1^{+0.8}_{-0.5} \times 10^6\,\Msun$
is inferred; unfortunately however, the extremely large uncertainty 
in the \4686\ lag renders the virial mass obtained from it
not very enlightening, but it is consistent with the \Hbeta\ result.

It is also important to keep in mind that because of the unknown
geometry and kinematics of the BLR, the virial mass is reliable
to only about an order of magnitude, i.e., the systematic uncertainties
are much larger than the errors quoted here. However, an independent
estimate of the inclination of the system has been made by
Christopoulou et al.\ (1997) on the basis of the NLR
morphology and kinematics. These authors model the NLR as
an outflowing biconical region of inclination 50\deg\ and
half-opening angle 23\deg. If the BLR and NLR axes are coaligned,
then correcting the virial mass for inclination gives
a central mass of 
$1.4^{+1.0}_{-0.6} \times 10^6\,\Msun$.

There are a number of important features in the rms
spectrum that deserve attention: first, the constant components, such as the 
[\oiii]\,$\lambda\lambda4959$, 5007 narrow lines that are so
prominent in the mean spectrum, are absent, as expected, in the rms spectrum
(except for weak residuals which reflect the accuracy to
which accurate flux calibration can be achieved). Second, the
weak broad wings of \Hbeta\ that can be seen in the mean spectrum
are much weaker in the rms spectrum, i.e., the line core is more
variable than the line wings. This could occur if the higher velocity
material is much farther away from the ionizing source, or if
some significant component of the high-velocity gas is optically
thin (e.g., Shields, Ferland, \& Peterson 1995); 
on physical grounds, we prefer the latter explanation. Third, 
the rms spectrum also shows that the optical \feii\
emission varied little, if at all, during this period.
The optical spectra of AGNs have broad blended \feii\
features that extend from just longward of \Hgamma\ to
just shortward of \Hbeta, and over the range $\sim5100$--5600\,\AA.
In the mean spectrum, these \feii\ emission features are quite strong, which
makes it hard to isolate the \4686\ emission. The \feii\
blends are, however, virtually absent in the rms spectrum.
Fourth, \4686\ is very prominent in the rms spectrum and is much 
broader than \Hbeta. The rest-frame width
of this line in the rms spectrum is $\sim5430$\,\kms, very 
typical of the line widths seen in normal Seyfert 1 galaxies.
The centroid of the \4686\ is strongly
{\em blueshifted} relative to \Hbeta, by about 1400\,\kms.
In order to demonstrate that these properties of the
\4686\ profile are real, we also computed the mean and rms
spectra based on (a) set A spectra obtained at the same time
and (b) set B spectra obtained during the 1997 monitoring season
(Year 2). The rms spectra from these subsets, shown in the
bottom two panels of Fig.\ 5, show the same \4686\ profile
characteristics seen in the set B Subset 1 data shown in
the second panel.

\section{Discussion}

\subsection{The Continuum}

As described in the previous section, the X-ray and optical
continuum variations are not closely coupled on short time scales.
The X-ray continuum shows large scale variations on short
time scales, as is typical of the NLS1 class. The rapid
variations seen in the X-ray are not detected in the optical,
as has previously been reported by Done et al.\ (1990) for this
same source and by Young et al.\ (1999) for IRAS 13224$-$3809.
However, if we average over the short time-scale flares, 
then the X-ray and optical continuum variations
{\em do} seem to be coupled, though the time resolution of this
experiment is insufficient to determine whether there is any
lag between them at the level of days or less. The absence of
strong coupling of the hard X-ray and optical
continuum variations argues against reprocessing models
in which hard X-rays are absorbed by a dense plasma (such as
the accretion disk) and the energy is re-radiated at longer
wavelengths (Guilbert \& Rees 1988). The X-ray variations
are more suggestive of localized flaring types of activity that
may arise in a patchy corona above the accretion disk.

It has already been pointed out based on these same {\em RXTE} data 
(Uttley et al.\ 1999) that the X-ray continuum
of NGC~4051 virtually ``turned off'' in early 1998 
(around JD 2450800; see Fig.\ 1). However, the optical spectroscopic
data show that the optical continuum and emission lines
(and therefore, by inference, the ionizing UV continuum)
did not disappear at the same time. This seems in a sense
rather suggestive of the kind of behavior that has
been seen in Galactic black-hole systems such as GRO J1655$-$40 
(Orosz et al.\ 1997).  A possible interpretation of 
the behavior of NGC~4051 is that the inner X-ray producing
(though not necessarily by purely thermal emission)
part of the accretion disk 
in NGC~4051 has entered
an advection-dominated accretion-flow (ADAF) state, in
which radiation is emitted with very low efficiency
(Narayan \& Yi 1994; Narayan et al.\ 1998). 
This implies that there is a transition radius
$r_{\rm tran}$ inside of which the disk is an ADAF and
outside of which it radiates efficiently, perhaps like
a classical thin disk (Shakura \& Sunyaev 1973). The
persistence of the optical continuum and emission lines suggests
that this transition radius is somewhere between the
regions that are most responsible for the soft X-rays and
the H-ionizing continuum. We comment on this further 
in section 4.3 below.

\subsection{The Virial Mass and Implications for NLS1s}

As noted above, reverberation-based size estimates for the
broad emission lines and resulting virial mass estimates
provide a potential means of distinguishing among the various
NLS1 models. In Fig.\ 6, we show the relationship between
the BLR radius as measured from the \Hbeta\ lag as a function
of the optical continuum luminosity for all AGNs with Balmer
line lags known to reasonable accuracy. All data used here
are taken from the compilation of Kaspi et al.\ (2000), though
their parameters for NGC 4051 are superceded by
the values reported here. This compilation contains six
additional AGNs that could in principle be classified as NLS1s
as they meet the criterion $\vFWHM \ltsim 2000$\,\kms. These
sources, which we shall refer to below as ``narrow-line objects'',
are the Seyfert galaxies Mrk~335 and Mrk~110 (from Wandel,
Peterson, \& Malkan 1999) and the QSOs
PG~0026$+$120, PG~1211$+$143, PG~1351$+$640, and
PG~1704$+$608 (from Kaspi et al.\ 2000).

The best-fit regression line ($R_{\rm BLR} \propto L^{0.62 \pm 0.02}$),
based on all objects from Table 5 of Kaspi et al.\ (2000)
except NGC~4051, is shown as a dotted line.
NGC~4051 lies approximately 2.8$\sigma$ above this regression
line, although all the other narrow-line objects clearly fall
in the locus defined by the AGNs with broader lines. It is 
difficult to argue that NGC~4051 is somehow different from 
the other AGNs, as there are several AGNs that have large
displacements from the regression line. This is reflected
in the large value of the reduced $\chi^2$
statistic, $\chi^2_{\nu} = 15.7$, for this fit. 

In Fig.\ 7, we plot the mass--luminosity relationship for the
AGNs from Kaspi et al., and we show (a) the best-fit regression line
based on all objects {\em except} the seven narrow-line objects
and (b) that based on the narrow-line objects alone. Formally, the slopes 
$\alpha$ (for $M \propto L^{\alpha}$) are significantly different,
with $\alpha = 0.46 \pm 0.06$ for the narrow-line objects 
(including NGC~4051), and $\alpha = 0.27 \pm 0.03$ for the others. 
These two fits are separated by typically an order of magnitude
in black-hole mass; the black holes in the narrow-line objects 
are about a factor of 10 lower than those of other AGNs of comparable
luminosity.

How well do these results allow us to distinguish among the
various explanations for the NLS1 phenomenon? We consider
the possibilities:
\begin{enumerate}
\item {\em Do the BLRs of NLS1s have anomalously large radii?}
The position of NGC~4051 in Fig.\ 6 might suggest that this is possible,
but the distribution of other narrow-line objects does not 
support this. Furthermore, as noted above, the scatter in
the BLR-radius luminosity relationship is very large, and 
NGC~4051 is in a statistical sense not the largest outlier in
this relationship (simply because other sources have smaller
uncertainties in their measured lags). 

\item {\em Are NLS1s simply low-inclination systems?}
If the BLR is a flattened system, at low inclination
(i.e., nearly face-on) the line widths will be decreased
by a factor $\sin i$, but the measured emission-line lags
will be relatively unaffected. On the other hand, assuming
that the UV--optical continuum arises in an accretion
disk at the same inclination, the apparent UV--optical luminosity
is higher at lower inclination (e.g., see Fig.\ 32
of Netzer 1990). Thus, relative to similar sources at
intermediate inclination, the masses of low-inclination
sources will be underestimated, and their luminosities will
be overestimated, displacing the narrow-line objects in
Fig.\ 7 towards the lower right. This is generally consistent
with the location of all seven of the narrow-line objects,
including NGC~4051. The line transfer function for \Hbeta\
would provide a  more definitive test of this hypothesis
since it would allow determination of the inclination of the system.
This would require more and better data than we have
obtained in this experiment.

However, as noted earlier, Christopoulou et al.\ (1997)
show that the NLR morphology and kinematics suggest a system
that is inclined to the line of sight by $\sim50$\deg.
It seems reasonable to suppose that the NLR and BLR axes
are approximately coaligned. If this is the case, then
our virial estimate for the central mass is too small by
a modest factor $\sin 50\deg \approx 0.77$, and the corrected
central mass is then
$1.4^{+1.0}_{-0.6} \times 10^6\,\Msun$.

\item {\em Are NLS1s undermassive systems with relatively 
high accretion rates?} Again, the distribution of the
narrow-line objects, including NGC~4051, in Fig.\ 7 is consistent
with this hypothesis. The narrow-line sources on this plot
lie below the mass-luminosity relationship for other AGNs,
at the lower end of the envelope around this relationship.

\end{enumerate}

In summary, the hypothesis that NLS1s have unusually distant BLRs
for their luminosity
is probably not viable in general, although it could apply to
the specific case of NGC~4051. At the present time, however,
we cannot distinguish between the low-inclination and
low-mass, high accretion-rate hypotheses on the basis of the
reverberation results alone. The latter is favored
on the basis X-ray considerations and 
the 50\deg\ inclination inferred from the NLR.
Indeed, it is entirely
possible that both effects (i.e., low inclination and low
black-hole mass) contribute. An improvement in 
the optical spectroscopic monitoring data 
could allow determination of the \Hbeta\ transfer function,
which could allow discrimination between these competing models.

\subsection{Broad He\,{\small\bf II} Emission}

As noted in section 3.2, we detected very broad,
blueshifted \4686\ emission in the rms spectra.
The blueshift of this
component is suggestive of radial rather than virialized
motion. There is no similar
obvious component to the Balmer lines, but we should expect
that similar blueshifted features might appear in the other
higher-ionization lines in the UV. As no contemporaneous
spectra were available, we retrieved the 31
{\em International Ultraviolet Explorer} 
SWP (Short-Wavelength Prime camera, wavelength range
$\sim1150$--2000\,\AA) spectra from the
Space Telescope Science Institute Multimission Data Archive. We formed an
average of all these spectra, since the individual
spectra were of rather low signal-to-noise ratio.
In Fig.\ 8, we show the line profiles of \4686\
(the rms profile, as in Fig.\ 5) and those of \heii\,$\lambda1640$ and
\civ\,$\lambda1549$ based on the mean {\em IUE} spectra. We note
that each of these lines has a strong wing extending several
thousand kilometers per second blueward of the line peak;
indeed, the comparatively large widths 
and blueshifts of the UV lines in NLS1
spectra has been noted earlier by Rodr\'{\i}guez-Pascual, 
Mas--Hesse, \& Santos-Lle\'{o} (1997).
It is possible that this gas is related to the known warm
absorber in NGC~4051 (McHardy et al.\ 1995). Interestingly,
photoionization equilibrium modeling of the X-ray warm absorber
data (Nicastro et al.\ 1999)
suggests a distance from the source of approximately
5 light days, which is consistent with the
reverberation result given in Table 4. 

The differences between the characteristics of the \Hbeta\ emission line
on one hand and of the high ionization lines on the other suggests a
two-component BLR, which has been proposed on numerous occasions on
other grounds (e.g., Collin-Souffrin et al.\ 1988;
van Groningen 1987).
In this particular case, an interpretation that is at least qualitatively
consistent with all the data and relatively simple
is that the Balmer lines arise primarily
in material that is in a flattened disk-like configuration at a low
to moderate 
inclination (to account for the narrow width of the \Hbeta\ line),
and the high-ionization lines arise in an outflowing wind,
of which we see preferentially the component on the near side of the disk
(to account for high velocity and blueward asymmetry).
Such a model is illustrated schematically by 
Collin-Souffrin et al.\ (1988; their Fig.\ 1) and more recently by
Dultzin-Hacyan, Taniguichi, \& Uranga (1999; their Fig.\ 1).
This geometry was also suggested by van Groningen (1987) to
explain the line profiles and profile ratios of the Balmer
lines in Mrk 335, another narrow-line object.

As noted earlier, our virial mass estimate of $M=1.1\times10^6$\,\Msun\
might seriously underestimate the black hole mass if the inclination
of the system is very low. Moreover,
some low-inclination accretion-disk models predict relatively
strong, variable EUV/soft X-ray fluxes (e.g., Netzer 1987; Madau 1988),
consistent with observations of NLS1s. We therefore cannot
rule out the possibility that the NLS1 phenomenon is due
principally to inclination effects. However, the 
strong rapid X-ray variability of NLS1s seem to favor the
low-mass, high accretion-rate explanation, as does
the 50\deg\ inclination of the NLR, unless the BLR and NLR axes of
symmetry are very different, which seems rather unlikely
on physical grounds.

The behavior of the \4686\ line during Year 3 may provide additional
information about the ionizing continuum at the time the X-rays
went into an extremely faint state. 
In Fig.\ 9, we show the rms spectrum based on
set B spectra obtained between JD2450810 and JD2451022. The \Hbeta\
line is strong and narrow, as it is in the other rms spectra shown
in Fig.\ 5, which indicates that the continuum shortward of 
912\,\AA\ is still present and variable. However, the 
\4686\ line is absent or very weak, indicating that the
driving extreme ultraviolet (EUV)
continuum shortward of the \heii\ edge at 228\,\AA\
(54.4\,eV) is {\em not} driving \4686\ variations, either because
the EUV flux is low or varying little. Evidence from earlier
years (Uttley et al.\ 1999) shows that the EUV and X-ray
fluxes in NGC~4051 are well correlated, which implies that the EUV continuum
might also have been extremely weak in Year 3. 
Simultaneous {\it BeppoSAX}, {\it RXTE}, and {\it EUVE}
observations obtained during the low state indicate that
this is correct (Uttley et al.\ 2000). If the
\heii\ emission has dramatically decreased during Year 3,
we could infer that the transition radius between the inner ADAF region
and the outer thin-disk structure must occur somewhere outside
(or in the vicinity of)
the region of the disk that contributes most of the flux
at about 228\,\AA, but inside the radius that contributes most
of the flux at about 912\,\AA. 

In this regard, whether or not there is residual
\4686\ emission in the low state during Year 3 becomes an interesting
question. Note in particular that the measurements of 
the \4686\ flux given in Table 2 and Figure 1 tell us little 
because these values include flux from
broad-line \feii\ blends and various narrow-line features. 
In order to address the question of how much \4686\ persists in
the low state, we have attempted a decomposition of the \4686\
feature. Our first step was to form average high-state (as in the top
panel of Fig.\ 5) and low-state spectra. We then attempted to
remove the \feii\ blends from the spectra by using the optical \feii\
template spectrum of Boroson \& Green (1982), convolved with a
Gaussian to match the width of the features in the spectrum of NGC~4051.
After subtraction of the flux-scaled and broadened \feii\ template,
we subtracted a power-law continuum from each spectrum. The 
\4686\ region of the resulting high-state and low-state spectra 
are shown in the top panel of Fig.\ 10. We then subtracted the
low-state spectrum from the high-state spectrum, forming the
difference spectrum shown in the middle panel of Fig.\ 10. The flux
in the \4686\ line in the difference spectrum is
$7.4 \times 10^{-15}$\,\lineunits. We
then make the assumption that the \heii\ line profile in the
difference spectrum can be used to model the \heii\ contribution
in the low-state spectrum. We then scale the \heii\ profile 
in flux on a trial-and-error basis and subtract it from the
low-state spectrum until the flux in the model profile causes
the residual to have negative flux. Using this procedure, we find
the largest possible value for the \4686\ line in the low-state is
$\sim8.8\times10^{-15}$\,\lineunits. These values mean that the
\4686\ flux decreased by at least 46\% between the Year 1
high state and the Year 3 low state. Our conclusion is
thus that while broad \4686\ probably did not completely disappear
during the low state in Year 3, the line flux did decrease
dramatically and any \heii\ variations during this period were
too small to detect.

In any case, these results underscore the importance of
multiwavelength observations of NLS1s in very low states.
Presumably comparison of the strength of the various
UV emission lines in the low state relative to the
values in the high state could lead to determination
of $r_{\rm tran}$. Unfortunately, no ultraviolet data are
available during the present campaign.

\section{Summary}
On the basis of three years of combined X-ray and optical spectroscopic
monitoring of the narrow-line Seyfert 1 galaxy NGC 4051, we
reach the following conclusions:
\begin{enumerate}
\item The rapid and strong X-ray variations that characterize narrow-line
Seyfert 1 galaxies are detected in our X-ray observations of
NGC~4051, but are not detected in the optical, consistent with
previous findings.
\item On time scales of many weeks and longer, there does appear to be
a correlation between the X-ray and optical continuum fluxes.
\item The variable part of the \Hbeta\ emission line has a Doppler
width of $\sim1100$\,\kms, and a time delay relative to the continuum
of about six days. Combining these quantities leads to a virial mass
estimate of $\sim1.1\times10^6$\,\Msun. If we assume that the
inclination of the system is 50\deg, as suggested by the NLR study
of Christopoulous et al.\ (1997), then the mass of the central
source is $\sim1.4\times10^6$\,\Msun. 
\item The \4686\ emission line is strongly variable, although an accurate
time delay cannot be measured. This line is about five times as broad
as the \Hbeta\ line, and is strongly blueward asymmetric, as are the
UV high-ionization lines in this object.
\item In the BLR radius--luminosity relationship, we find that narrow-line
objects (those with $\vFWHM \ltsim 2000$\,\kms) seem to fall on the same
locus as AGNs with broad lines.
\item In the virial mass--luminosity relationship, narrow-line objects
populate the low-mass end of a rather broad envelope; they have
virial masses typically an order of magnitude lower than other AGNs of
similar luminosity.
\item During the third year of this program, the hard X-ray flux decreased
by approximately a factor of 10, and the broad-line \4686\ flux
decrease by nearly a factor of two and did not show detectable
variations during this low state. At the same time,
the optical continuum and broad \Hbeta\
emission line were only slightly fainter than previously and continued
to vary significantly. This suggests that the innermost part of the
accretion disk went into an ADAF state, greatly reducing the production
of high-energy continuum photons from the inner part of the accretion
disk. 
\end{enumerate}

A picture that is consistent with the emission-line characteristics is
one in which the Balmer lines arise primarily in a disk-like configuration seen
at low-to-moderate inclination and the high-ionization lines arise 
primarily in an outflowing 
wind. The high-ionization lines are blueward asymmetric because we see 
emission preferentially from the near side, with the far side 
at least partially obscured by the
disk component (which might be an extension of the accretion disk).
This geometry requires that the much of the high-ionization
line flux arises in a region of scale comparable to the 
disk system that emits the low-ionization lines and that 
the disk system is at least partially opaque to the 
\heii\ line radiation. 
If NGC 4051 is typical of the NLS1 class, 
then it might be that NLS1s are best described 
as low-mass, high accretion-rate systems, although
the possible role of inclination cannot be discounted.
Indeed, the full explanation of the NLS1 phenomenon may involve {\em both}
inclination and black-hole mass.

\bigskip
For support of this work, we are grateful to 
the National Science Foundation (grant AST--9420080 
to The Ohio State University). We thank the referee, M.-H.\ Ulrich,
for suggestions that improved this paper.

\newpage



\clearpage

\begin{figure}
\caption{Continuum and emission-line light curves for NGC 4051 between
1996 January and 1998 June.
The top panel shows the X-ray light curve for NGC~4051 in
counts per second over the 2--10\,keV bandpass;
one count per second corresponds to a flux of approximately
$10^{-12}$ ergs cm$^{-2}$ s$^{-1}$ over the 2--10\,keV range.
The second panel shows the optical continuum light
curve in units of  $10^{-15}$\,\contunits, as in Table 2.
The third and fourth panels show, respectively, the
\Hbeta\ and \4686\ emission-line light curves, in 
units of  $10^{-13}$\,\lineunits, also as in Table 2. }
\end{figure}

\begin{figure}
\caption{The light curves in Fig.\ 1 are expanded to show
the well-sampled region used in the optical continuum/
emission-line cross-correlation analysis (``subset1'')
between JD 2450183 and  JD 2450262. Units are as in
Fig.\ 1.}
\end{figure}

\begin{figure}
\caption{The mean X-ray (filled circles, flux scale on left axis) 
and optical continuum (open circles, flux scale on right axis) light curves 
from as in Fig.\ 1 smoothed with a rectangular function of
width 30 days. This shows when the 
rapid X-ray variability is suppressed by averaging, the X-ray
and optical variations are correlated on time scales of several weeks
and longer.}
\end{figure}

\begin{figure}
\caption{Emission-line cross-correlation functions are
shown, interpolation cross-correlation function as solid
lines and discrete correlation function values as points
with associated error bars.  
The top panel shows the 5100\,\AA\ continuum--\Hbeta\
cross-correlation, along with the continuum autocorrelation
function shown as a dotted line. The bottom panel
shows the 5100\,\AA\ continuum--\4686\ cross-correlation.
Peak and centroid values are given in Table 4.}
\end{figure}

\begin{figure}
\caption{Mean (top panel) and root-mean-square (second panel)
spectra of NGC~4051 based on individual spectra  obtained
during the well-sampled interval between JD 2450183 and
JD 2450262 (``subset 1''), as in Fig.\ 2.
The spectra are plotted in the observed frame, and
in units of $10^{-15}$\,\contunits, as in Table 2.
These are based only on the data from set B. The
\4686\ line is very broad (${\rm FWHM} \approx 5430$\,\kms) and
blueshifted (by $\sim 1400$\,\kms\ relative to the narrow
emission lines); these characteristics are also seen
in the rms spectrum based on contemporaneous spectra
from set A (third panel) and on set B spectra obtained
the next year (bottom panel).}
\end{figure}

\begin{figure}
\caption{The relationship between the size of the broad
Balmer-line emitting region and optical luminosity for
AGNs. The filled circles are Seyfert galaxies from
Wandel, Peterson, \& Malkan (1999) and the open circles are
QSOs from Kaspi et al.\ (2000). The large triangles those
AGNs from the same sources whose broad lines have widths
less than $\sim 2000$\,\kms, i.e., the Seyfert galaxies
Mrk~335 and Mrk~110 and the QSOs
PG~0026$+$120, PG~1211$+$143, PG~1351$+$640, and
PG~1704$+$608. NGC~4051 is shown as a filled diamond.}
\end{figure}

\begin{figure}
\caption{The relationship between the reverberation-based
virial mass and  and optical luminosity for
AGNs. Date points are coded as in Fig.\  6.}
\end{figure}

\begin{figure}
\caption{The top panel shows the profile of \4686\ in
the rms spectrum from Year 1 (1996), as in Fig.\ 5.
The profiles of \heii\,$\lambda1640$ and \civ\,$\lambda1549$ from
mean {\em IUE} spectra are shown in the middle and
bottom panels, respectively. Fluxes are in
units of $10^{-15}$\,\contunits,
and in 	each panel the velocity scale is based on
an assumed systemic redshift of $z = 0.002418$ 
(deVaucouleurs et al.\ 1991). The blue wing appears more pronounced
in the top panel because the rms spectrum enhances the variable
part of the line. Also, the optical and UV observations
are {\em not} contemporaneous.}
\end{figure}

\begin{figure}
\caption{The rms spectrum based on the set B spectra obtained
during Year 3 between JD2450810 and JD2451022. Note that 
\Hbeta\ remains strongly variable, but in comparison to the
rms spectra shown in Fig.\ 5, the \4686\ line is much
weaker, if present at all.}
\end{figure}

\begin{figure}
\caption{The top panel shows the \4686\ region of the mean
high-state and low-state spectra, with \feii\ emission
and the underlying continuum subtracted. The middle panel
shows the difference between the high-state and low-state
spectra. The low-state spectrum is shown for reference
(note in particular the strong \4686\ narrow-line component
in the low-state spectrum. The bottom panel shows our
best estimate of the low-state spectrum with the largest
possible contribution of broad \heii\ emission removed.
Data are plotted in the rest frame of NGC~4051. The
vertical axis is flux in units of $10^{-15}$\,\contunits.}
\end{figure}

\clearpage

\begin{deluxetable}{lcccl} 
\tablewidth{0pt}
\tablecaption{Log of Spectroscopic Observations}
\tablehead{
\colhead{UT} &
\colhead{Julian Date} &
\colhead{ } &
\colhead{Range} &
\colhead{File} \nl
\colhead{Date} &
\colhead{$(-2450000)$} &
\colhead{Code} &
\colhead{(\AA)} &
\colhead{Name} \nl
\colhead{(1)} &
\colhead{(2)} &
\colhead{(3)} &
\colhead{(4)} &
\colhead{(5)} 
}
\startdata
1996 Jan 12 &  95.0 & B & 3670--7450 & n00095b\nl
1996 Jan 15 &  98.0 & B & 3650--7520 & n00098b\nl
1996 Jan 19 & 102.0 & B & 3650--7410 & n00102b\nl
1996 Jan 22 & 105.1 & B & 3660--7410 & n00105b\nl
1996 Jan 25 & 108.0 & B & 3650--7520 & n00108b\nl
1996 Jan 29 & 112.0 & B & 3640--7500 & n00112b\nl
1996 Feb 05 & 118.8 & A & 4510--5670 & n00118a\nl
1996 Feb 09 & 122.8 & A & 4520--5670 & n00122a\nl
1996 Feb 10 & 124.0 & B & 3660--7400 & n00123b\nl
1996 Feb 14 & 127.8 & A & 4520--5670 & n00127a\nl
1996 Feb 15 & 129.0 & B & 3660--7420 & n00129b\nl
1996 Feb 18 & 132.0 & B & 3660--7530 & n00132b\nl
1996 Feb 21 & 134.9 & B & 3660--7410 & n00134b\nl
1996 Feb 23 & 136.8 & A & 4520--5680 & n00136a\nl
1996 Mar 07 & 149.8 & A & 4520--5680 & n00149a\nl
1996 Mar 22 & 164.8 & A & 4530--5680 & n00164a\nl
1996 Mar 25 & 167.7 & B & 3640--7500 & n00167b\nl
1996 Mar 26 & 168.8 & A & 4520--5690 & n00168a\nl
1996 Mar 28 & 170.8 & B & 3650--7390 & n00170b\nl
1996 Apr 10 & 183.6 & B & 3660--7420 & n00183b\nl
1996 Apr 12 & 185.6 & B & 3670--7420 & n00185b\nl
1996 Apr 12 & 185.7 & A & 4540--5690 & n00185a\nl
1996 Apr 15 & 188.6 & B & 3660--7410 & n00188b\nl
1996 Apr 18 & 191.6 & B & 3660--7410 & n00191b\nl
1996 Apr 20 & 193.6 & B & 3660--7400 & n00193b\nl
1996 Apr 25 & 198.7 & A & 4510--5680 & n00198a\nl
1996 Apr 25 & 198.8 & B & 3650--7410 & n00198b\nl
1996 Apr 26 & 199.6 & B & 3660--7400 & n00199b\nl
1996 May 03 & 206.7 & A & 4530--5690 & n00206a\nl
1996 May 08 & 211.6 & B & 3660--7400 & n00211b\nl
1996 May 09 & 212.8 & A & 4540--5680 & n00212a\nl
1996 May 10 & 213.6 & B & 3660--7520 & n00213b\nl
1996 May 15 & 218.6 & B & 3660--7510 & n00218b\nl
1996 May 17 & 220.8 & A & 4540--5700 & n00220a\nl
1996 May 18 & 221.7 & B & 3660--7400 & n00221b\nl
1996 May 22 & 225.6 & B & 3670--7430 & n00225b\nl
1996 May 24 & 227.8 & A & 4540--5700 & n00227a\nl
1996 May 26 & 229.7 & B & 3670--7430 & n00229b\nl
1996 May 30 & 233.7 & A & 4530--5670 & n00233a\nl
1996 Jun 06 & 240.6 & B & 3660--7400 & n00240b\nl
1996 Jun 07 & 241.7 & A & 4530--5690 & n00241a\nl
1996 Jun 14 & 248.7 & A & 4540--5690 & n00248a\nl
1996 Jun 16 & 250.7 & B & 3600--7540 & n00250b\nl
1996 Jun 19 & 253.6 & B & 3670--7530 & n00253b\nl
1996 Jun 19 & 253.7 & A & 4520--5680 & n00253a\nl
1996 Jun 23 & 257.7 & B & 3660--7420 & n00257b\nl
1996 Jun 25 & 259.6 & B & 3670--7550 & n00259b\nl
1996 Jun 28 & 262.7 & A & 4520--5680 & n00262a\nl
1996 Jul 17 & 281.6 & B & 3670--7520 & n00281b\nl
1996 Jul 20 & 284.7 & B & 3670--7530 & n00284b\nl
1996 Jul 24 & 288.6 & B & 3670--7530 & n00288b\nl
1996 Dec 11 & 429.0 & B & 3650--7510 & n00429b\nl
1996 Dec 17 & 435.0 & B & 3660--7520 & n00435b\nl
1997 Jan 02 & 451.0 & B & 3670--7510 & n00451b\nl
1997 Jan 09 & 458.0 & B & 3670--7510 & n00458b\nl
1997 Jan 16 & 465.0 & B & 4000--7500 & n00465b\nl
1997 Jan 31 & 479.9 & A & 4720--5990 & n00479a\nl
1997 Jan 31 & 480.0 & B & 4000--7500 & n00480b\nl
1997 Feb 03 & 483.1 & B & 4000--7500 & n00483b\nl
1997 Feb 06 & 485.9 & B & 4000--7530 & n00485b\nl
1997 Feb 09 & 489.0 & B & 4000--7500 & n00488b\nl
1997 Feb 14 & 493.9 & A & 4380--5550 & n00493a\nl
1997 Feb 14 & 494.0 & B & 4000--7520 & n00494b\nl
1997 Feb 27 & 506.8 & A & 4360--5500 & n00506a\nl
1997 Mar 02 & 510.0 & B & 3660--7520 & n00509b\nl
1997 Mar 12 & 519.8 & B & 3660--7500 & n00519b\nl
1997 Mar 13 & 520.8 & A & 4290--5820 & n00520a\nl
1997 Mar 14 & 521.8 & B & 3670--7310 & n00521b\nl
1997 Mar 20 & 527.8 & A & 4390--5940 & n00527a\nl
1997 Apr 09 & 547.6 & B & 3670--7500 & n00547b\nl
1997 Apr 12 & 550.8 & B & 3670--7520 & n00550b\nl
1997 Apr 13 & 551.8 & B & 3680--7540 & n00551b\nl
1997 Apr 29 & 567.6 & B & 3670--7530 & n00567b\nl
1997 May 01 & 569.6 & B & 3670--7540 & n00569b\nl
1997 May 08 & 576.6 & B & 3650--7500 & n00576b\nl
1997 May 10 & 578.6 & B & 3650--7520 & n00578b\nl
1997 May 12 & 580.7 & B & 3660--7520 & n00580b\nl
1997 May 14 & 582.6 & B & 3650--7530 & n00582b\nl
1997 May 29 & 597.7 & B & 3660--7520 & n00597b\nl
1997 Jun 01 & 600.6 & B & 4000--7530 & n00600b\nl
1997 Jun 03 & 602.6 & B & 3660--7540 & n00602b\nl
1997 Jun 05 & 604.6 & B & 3670--7530 & n00604b\nl
1997 Jun 09 & 608.6 & B & 3670--7530 & n00608b\nl
1997 Jun 11 & 610.6 & B & 3670--7540 & n00610b\nl
1997 Jun 28 & 627.6 & B & 3670--7530 & n00627b\nl
1997 Jul 01 & 630.7 & B & 3670--7520 & n00630b\nl
1997 Jul 06 & 635.6 & B & 3670--7530 & n00635b\nl
1997 Jul 12 & 641.6 & B & 3670--7530 & n00641b\nl
1997 Jul 14 & 643.6 & B & 3670--7520 & n00643b\nl
1997 Nov 22 & 775.0 & A & 4310--5800 & n00775a\nl
1997 Nov 24 & 777.0 & B & 3750--7510 & n00777b\nl
1997 Nov 29 & 782.0 & B & 3750--7500 & n00782b\nl
1997 Dec 04 & 787.0 & B & 3750--7500 & n00787b\nl
1997 Dec 27 & 810.1 & B & 3670--7530 & n00810b\nl
1998 Jan 20 & 834.0 & B & 3680--7540 & n00834b\nl 
1998 Jan 24 & 838.1 & B & 3660--7550 & n00838b\nl 
1998 Jan 25 & 839.0 & A & 4330--5840 & n00839a\nl
1998 Jan 26 & 840.1 & B & 3670--7540 & n00840b\nl 
1998 Jan 29 & 843.0 & B & 3650--7520 & n00843b\nl 
1998 Feb 02 & 847.0 & B & 3660--7540 & n00847b\nl 
1998 Feb 22 & 867.0 & B & 3650--7530 & n00867b\nl 
1998 Feb 24 & 869.0 & B & 3670--7540 & n00869b\nl 
1998 Feb 28 & 873.0 & B & 3670--7530 & n00872b\nl 
1998 Mar 03 & 875.8 & A & 4300--5830 & n00876a\nl
1998 Mar 03 & 876.0 & B & 3660--7530 & n00876b\nl 
1998 Mar 04 & 877.0 & B & 3660--7510 & n00877b\nl 
1998 Mar 13 & 885.8 & A & 4380--5910 & n00886a\nl
1998 Mar 20 & 892.8 & B & 3660--7510 & n00892b\nl 
1998 Apr 03 & 906.9 & B & 3630--7490 & n00906b\nl 
1998 Apr 18 & 921.9 & B & 3610--7470 & n00921b\nl 
1998 Apr 23 & 926.8 & A & 4320--5840 & n00927a\nl
1998 Apr 27 & 930.8 & B & 3620--7500 & n00930b\nl 
1998 May 02 & 935.9 & B & 3630--7490 & n00935b\nl 
1998 May 03 & 936.6 & B & 3620--7490 & n00936b\nl 
1998 May 16 & 949.6 & B & 3670--7510 & n00949b\nl 
1998 May 28 & 961.6 & B & 3660--7530 & n00961b\nl 
1998 Jun 02 & 966.7 & B & 3660--7540 & n00966b\nl 
1998 Jun 16 & 980.7 & A & 4310--5850 & n00981a\nl
1998 Jun 16 & 980.7 & B & 3660--7540 & n00980b\nl 
1998 Jun 19 & 983.7 & B & 3650--7520 & n00983b\nl 
1998 Jun 24 & 988.6 & B & 3680--7530 & n00988b\nl 
1998 Jun 27 & 991.7 & B & 3660--7530 & n00991b\nl 
1998 Jun 30 & 994.6 & B & 3680--7520 & n00994b\nl 
1998 Jul 15 &1009.6 & B & 3710--7510 & n01009b\nl 
1998 Jul 18 &1012.6 & B & 3670--7530 & n01012b\nl 
1998 Jul 25 &1019.6 & B & 3700--7500 & n01019b\nl 
1998 Jul 28 &1022.6 & B & 3670--7530 & n01022b\nl 
\hline
\multicolumn{5}{c}{Codes for Data Origin }\nl
\multicolumn{1}{r}{A} &
\multicolumn{4}{l}{1.8-m Perkins Telescope + Ohio State CCD Spectrograph}\nl
\multicolumn{1}{r}{B} &
\multicolumn{4}{l}{1.5-m Tillinghast Reflector + FAST Spectrograph}
\enddata

\end{deluxetable}

\begin{deluxetable}{cccc} 
\tablewidth{0pt}
\tablecaption{Light Curves}
\tablehead{
\colhead{Julian Date} &
\colhead{$F_{\lambda}({\rm 5100\,\AA})$} &
\colhead{$F({\rm H}\beta)$} &
\colhead{$F({\mbox{He\,{\sc ii}\,$\lambda4686$}})$} \nl
\colhead{($-2450000$)} &
\colhead{($10^{-15}$\,ergs\,s$^{-1}$\,cm$^{-2}$\,\AA$^{-1}$)} &
\multicolumn{2}{c}{($10^{-13}$\,ergs\,s$^{-1}$\,cm$^{-2}$)} \nl
\colhead{(1)} &
\colhead{(2)} &
\colhead{(3)} &
\colhead{(4)} 
}
\startdata
   95.0 &$ 13.41 \pm  0.54$ & $  4.17 \pm  0.17$ & $  4.06 \pm  0.45$ \nl
   98.0 &$ 13.88 \pm  0.56$ & $  4.54 \pm  0.18$ & $  4.22 \pm  0.46$ \nl
  102.0 &$ 14.42 \pm  0.58$ & $  4.91 \pm  0.20$ & $  4.53 \pm  0.50$ \nl
  105.1 &$ 13.46 \pm  0.54 $ & $  4.54 \pm  0.18 $ &  $\ldots$ \nl
  108.0 &$ 13.75 \pm  0.55 $ & $  4.39 \pm  0.17 $ &  $\ldots$ \nl
  112.0 &$ 13.75 \pm  0.55$ & $  4.42 \pm  0.18$ & $  3.80 \pm  0.42$ \nl
  118.8 &$ 14.01 \pm  0.28 $ & $  4.20 \pm  0.08 $ &  $\ldots$ \nl
  122.8 &$ 14.12 \pm  0.28 $ & $  4.67 \pm  0.09 $ &  $\ldots$ \nl
  124.0 &$ 13.74 \pm  0.55$ & $  4.72 \pm  0.19$ & $  3.81 \pm  0.42$ \nl
  127.8 &$ 13.51 \pm  0.27 $ & $  4.66 \pm  0.09 $ &  $\ldots$ \nl
  129.0 &$ 14.37 \pm  0.57$ & $  5.09 \pm  0.20$ & $  3.97 \pm  0.44$ \nl
  132.0 &$ 14.91 \pm  0.60$ & $  4.42 \pm  0.18$ & $  3.88 \pm  0.43$ \nl
  134.9 &$ 13.39 \pm  0.54$ & $  4.63 \pm  0.19$ & $  3.99 \pm  0.44$ \nl
  136.8 &$ 14.13 \pm  0.28 $ & $  4.53 \pm  0.09 $ &  $\ldots$ \nl
  149.8 &$ 14.70 \pm  0.29 $ & $  4.91 \pm  0.10 $ &  $\ldots$ \nl
  164.8 &$ 14.29 \pm  0.29 $ & $  4.71 \pm  0.09 $ &  $\ldots$ \nl
  167.7 &$ 14.66 \pm  0.59$ & $  4.92 \pm  0.20$ & $  3.86 \pm  0.43$ \nl
  168.8 &$ 14.58 \pm  0.29 $ & $  4.75 \pm  0.09 $ &  $\ldots$ \nl
  170.8 &$ 14.17 \pm  0.57$ & $  5.06 \pm  0.20$ & $  4.14 \pm  0.46$ \nl
  183.6 &$ 14.66 \pm  0.59$ & $  4.84 \pm  0.19$ & $  4.53 \pm  0.50$ \nl
  185.6 &$ 14.51 \pm  0.58$ & $  5.29 \pm  0.21$ & $  3.86 \pm  0.42$ \nl
  185.7 &$ 14.06 \pm  0.28 $ & $  5.12 \pm  0.10 $ &  $\ldots$ \nl
  188.6 &$ 14.61 \pm  0.58$ & $  5.17 \pm  0.21$ & $  4.73 \pm  0.52$ \nl
  191.6 &$ 14.63 \pm  0.58$ & $  5.35 \pm  0.21$ & $  4.78 \pm  0.52$ \nl
  193.6 &$ 15.02 \pm  0.60 $ & $  5.75 \pm  0.23 $ &  $\ldots$ \nl
  198.7 &$ 14.43 \pm  0.29 $ & $  5.63 \pm  0.11 $ &  $\ldots$ \nl
  198.8 &$ 13.54 \pm  0.54$ & $  5.55 \pm  0.22$ & $  4.47 \pm  0.49$ \nl
  199.6 &$ 14.02 \pm  0.56$ & $  5.54 \pm  0.22$ & $  4.83 \pm  0.53$ \nl
  206.7 &$ 14.08 \pm  0.28 $ & $  5.63 \pm  0.11 $ &  $\ldots$ \nl
  211.6 &$ 13.54 \pm  0.54$ & $  5.48 \pm  0.22$ & $  4.63 \pm  0.51$ \nl
  212.8 &$ 13.15 \pm  0.26 $ & $  5.36 \pm  0.11 $ &  $\ldots$ \nl
  213.6 &$ 13.16 \pm  0.53$ & $  5.28 \pm  0.21$ & $  4.73 \pm  0.52$ \nl
  218.6 &$ 12.95 \pm  0.52$ & $  5.09 \pm  0.20$ & $  4.28 \pm  0.47$ \nl
  220.8 &$ 12.27 \pm  0.25 $ & $  4.59 \pm  0.09 $ &  $\ldots$ \nl
  221.7 &$ 13.34 \pm  0.53$ & $  4.28 \pm  0.17$ & $  3.08 \pm  0.34$ \nl
  225.6 &$ 12.26 \pm  0.49$ & $  4.37 \pm  0.17$ & $  2.91 \pm  0.32$ \nl
  227.8 &$ 13.14 \pm  0.26 $ & $  4.14 \pm  0.08 $ &  $\ldots$ \nl
  229.7 &$ 13.91 \pm  0.56$ & $  4.30 \pm  0.17$ & $  3.77 \pm  0.41$ \nl
  233.7 &$ 12.84 \pm  0.26 $ & $  4.51 \pm  0.09 $ &  $\ldots$ \nl
  240.6 &$ 13.18 \pm  0.53$ & $  4.88 \pm  0.19$ & $  3.76 \pm  0.41$ \nl
  241.7 &$ 13.29 \pm  0.27 $ & $  4.78 \pm  0.10 $ &  $\ldots$ \nl
  248.7 &$ 13.93 \pm  0.28 $ & $  5.64 \pm  0.11 $ &  $\ldots$ \nl
  250.7 &$ 12.99 \pm  0.52$ & $  5.68 \pm  0.23$ & $  4.59 \pm  0.50$ \nl
  253.6 &$ 11.67 \pm  0.47$ & $  5.22 \pm  0.21$ & $  3.53 \pm  0.39$ \nl
  253.7 &$ 12.52 \pm  0.25 $ & $  4.94 \pm  0.10 $ &  $\ldots$ \nl
  257.7 &$ 12.09 \pm  0.48$ & $  4.47 \pm  0.18$ & $  3.45 \pm  0.38$ \nl
  259.6 &$ 11.92 \pm  0.48$ & $  4.57 \pm  0.18$ & $  3.35 \pm  0.37$ \nl
  262.7 &$ 12.41 \pm  0.25 $ & $  4.17 \pm  0.08 $ &  $\ldots$ \nl
  281.6 &$ 13.64 \pm  0.55$ & $  5.25 \pm  0.21$ & $  4.17 \pm  0.46$ \nl
  284.7 &$ 14.68 \pm  0.59 $ & $  5.89 \pm  0.24 $ &  $\ldots$ \nl
  288.6 &$ 13.25 \pm  0.53$ & $  5.45 \pm  0.22$ & $  4.74 \pm  0.52$ \nl
  429.0 &$ 12.98 \pm  0.52$ & $  5.84 \pm  0.23$ & $  4.46 \pm  0.49$ \nl
  435.0 &$ 13.53 \pm  0.54$ & $  5.46 \pm  0.22$ & $  4.71 \pm  0.52$ \nl
  451.0 &$ 12.84 \pm  0.51$ & $  5.51 \pm  0.22$ & $  4.05 \pm  0.45$ \nl
  458.0 &$ 13.25 \pm  0.53$ & $  5.38 \pm  0.22$ & $  4.34 \pm  0.48$ \nl
  465.0 &$ 12.30 \pm  0.49$ & $  4.74 \pm  0.19$ & $  4.56 \pm  0.50$ \nl
  479.9 &$ 12.81 \pm  0.26 $ & $  5.04 \pm  0.10 $ &  $\ldots$ \nl
  480.0 &$ 12.68 \pm  0.51$ & $  5.10 \pm  0.20$ & $  4.67 \pm  0.51$ \nl
  483.1 &$ 11.95 \pm  0.48$ & $  4.69 \pm  0.19$ & $  3.08 \pm  0.34$ \nl
  485.9 &$ 12.41 \pm  0.50$ & $  4.70 \pm  0.19$ & $  3.22 \pm  0.35$ \nl
  489.0 &$ 14.03 \pm  0.56$ & $  4.92 \pm  0.20$ & $  3.99 \pm  0.44$ \nl
  493.9 &$ 12.67 \pm  0.25 $ & $  4.99 \pm  0.10 $ &  $\ldots$ \nl
  494.0 &$ 13.53 \pm  0.54$ & $  4.83 \pm  0.19$ & $  3.30 \pm  0.36$ \nl
  506.8 &$ 11.46 \pm  0.23 $ & $  4.69 \pm  0.09 $ &  $\ldots$ \nl
  510.0 &$ 12.30 \pm  0.49$ & $  4.37 \pm  0.17$ & $  3.27 \pm  0.36$ \nl
  519.8 &$ 12.51 \pm  0.50$ & $  4.74 \pm  0.19$ & $  3.68 \pm  0.41$ \nl
  520.8 &$ 12.50 \pm  0.25 $ & $  5.12 \pm  0.10 $ &  $\ldots$ \nl
  521.8 &$ 12.65 \pm  0.51$ & $  4.68 \pm  0.19$ & $  3.49 \pm  0.38$ \nl
  527.8 &$ 10.92 \pm  0.22 $ & $  4.57 \pm  0.09 $ &  $\ldots$ \nl
  547.6 &$ 11.80 \pm  0.47$ & $  4.52 \pm  0.18$ & $  3.57 \pm  0.39$ \nl
  550.8 &$ 11.74 \pm  0.47$ & $  4.47 \pm  0.18$ & $  2.96 \pm  0.32$ \nl
  551.8 &$ 11.59 \pm  0.46$ & $  4.42 \pm  0.18$ & $  3.11 \pm  0.34$ \nl
  567.6 &$ 10.88 \pm  0.44$ & $  4.48 \pm  0.18$ & $  3.76 \pm  0.41$ \nl
  569.6 &$ 11.76 \pm  0.47$ & $  4.31 \pm  0.17$ & $  3.12 \pm  0.34$ \nl
  576.6 &$ 12.17 \pm  0.49$ & $  4.12 \pm  0.17$ & $  2.91 \pm  0.32$ \nl
  578.6 &$ 12.02 \pm  0.48$ & $  4.32 \pm  0.17$ & $  3.24 \pm  0.36$ \nl
  580.7 &$ 11.97 \pm  0.48$ & $  4.91 \pm  0.20$ & $  3.28 \pm  0.36$ \nl
  582.6 &$ 12.10 \pm  0.48$ & $  4.49 \pm  0.18$ & $  3.35 \pm  0.37$ \nl
  597.7 &$ 12.57 \pm  0.50$ & $  4.24 \pm  0.17$ & $  2.92 \pm  0.32$ \nl
  600.6 &$ 12.44 \pm  0.50$ & $  4.15 \pm  0.17$ & $  2.59 \pm  0.28$ \nl
  602.6 &$ 11.62 \pm  0.47$ & $  4.61 \pm  0.18$ & $  3.53 \pm  0.39$ \nl
  604.6 &$ 12.00 \pm  0.48$ & $  4.50 \pm  0.18$ & $  3.59 \pm  0.40$ \nl
  608.6 &$ 11.96 \pm  0.48$ & $  4.16 \pm  0.17$ & $  3.39 \pm  0.37$ \nl
  610.6 &$ 10.99 \pm  0.44$ & $  4.24 \pm  0.17$ & $  2.80 \pm  0.31$ \nl
  627.6 &$ 11.26 \pm  0.45$ & $  4.01 \pm  0.16$ & $  2.92 \pm  0.32$ \nl
  630.7 &$ 11.14 \pm  0.45$ & $  3.88 \pm  0.16$ & $  2.49 \pm  0.27$ \nl
  635.6 &$ 11.87 \pm  0.47$ & $  4.24 \pm  0.17$ & $  2.37 \pm  0.26$ \nl
  641.6 &$ 12.33 \pm  0.49$ & $  4.71 \pm  0.19$ & $  3.20 \pm  0.35$ \nl
  643.6 &$ 10.75 \pm  0.43$ & $  4.46 \pm  0.18$ & $  2.83 \pm  0.31$ \nl
  775.0 &$ 13.29 \pm  0.27 $ & $  5.04 \pm  0.10 $ &  $\ldots$ \nl
  777.0 &$ 12.69 \pm  0.51$ & $  5.18 \pm  0.21$ & $  3.61 \pm  0.40$ \nl
  782.0 &$ 12.86 \pm  0.51$ & $  5.41 \pm  0.22$ & $  3.83 \pm  0.42$ \nl
  787.0 &$ 13.49 \pm  0.54$ & $  5.27 \pm  0.21$ & $  4.20 \pm  0.46$ \nl
  810.1 &$ 14.06 \pm  0.56$ & $  3.95 \pm  0.16$ & $  2.97 \pm  0.33$ \nl
  834.0 &$ 12.00 \pm  0.48$ & $  4.85 \pm  0.19$ & $  2.87 \pm  0.31$ \nl
  838.1 &$ 11.44 \pm  0.46$ & $  4.59 \pm  0.18$ & $  2.99 \pm  0.33$ \nl
  839.0 &$ 11.89 \pm  0.24 $ & $  4.89 \pm  0.10 $ &  $\ldots$ \nl
  840.1 &$ 12.50 \pm  0.50$ & $  4.91 \pm  0.20$ & $  3.40 \pm  0.37$ \nl
  843.0 &$ 12.45 \pm  0.50$ & $  4.87 \pm  0.19$ & $  3.29 \pm  0.36$ \nl
  847.0 &$ 12.33 \pm  0.49$ & $  4.82 \pm  0.19$ & $  3.20 \pm  0.35$ \nl
  867.0 &$ 12.25 \pm  0.49$ & $  4.20 \pm  0.17$ & $  2.86 \pm  0.31$ \nl
  869.0 &$ 12.55 \pm  0.50$ & $  3.85 \pm  0.15$ & $  2.53 \pm  0.28$ \nl
  873.0 &$ 12.15 \pm  0.49$ & $  4.09 \pm  0.16$ & $  3.01 \pm  0.33$ \nl
  875.8 &$ 12.29 \pm  0.25 $ & $  4.49 \pm  0.09 $ &  $\ldots$ \nl
  876.0 &$ 12.58 \pm  0.50$ & $  4.31 \pm  0.17$ & $  3.09 \pm  0.34$ \nl
  877.0 &$ 12.44 \pm  0.50$ & $  4.37 \pm  0.17$ & $  3.09 \pm  0.34$ \nl
  885.8 &$ 11.54 \pm  0.23 $ & $  4.18 \pm  0.08 $ &  $\ldots$ \nl
  892.8 &$ 12.14 \pm  0.49$ & $  4.54 \pm  0.18$ & $  3.05 \pm  0.34$ \nl
  906.9 &$ 11.44 \pm  0.46$ & $  4.53 \pm  0.18$ & $  2.60 \pm  0.28$ \nl
  921.9 &$ 11.61 \pm  0.46$ & $  4.53 \pm  0.18$ & $  2.93 \pm  0.32$ \nl
  926.8 &$ 11.13 \pm  0.22 $ & $  4.43 \pm  0.09 $ &  $\ldots$ \nl
  930.8 &$ 11.43 \pm  0.46$ & $  4.01 \pm  0.16$ & $  2.74 \pm  0.30$ \nl
  935.9 &$ 11.92 \pm  0.48 $ & $  4.51 \pm  0.18 $ &  $\ldots$ \nl
  936.6 &$ 11.23 \pm  0.45$ & $  4.39 \pm  0.18$ & $  2.65 \pm  0.29$ \nl
  949.6 &$ 12.03 \pm  0.48$ & $  5.04 \pm  0.20$ & $  3.41 \pm  0.38$ \nl
  961.6 &$ 11.32 \pm  0.45$ & $  4.90 \pm  0.20$ & $  3.09 \pm  0.34$ \nl
  966.7 &$ 11.60 \pm  0.46$ & $  4.39 \pm  0.18$ & $  3.00 \pm  0.33$ \nl
  980.7 &$ 11.72 \pm  0.21$ & $  4.73 \pm  0.08$ & $  3.13 \pm  0.34$ \nl
  983.7 &$ 11.41 \pm  0.46$ & $  4.46 \pm  0.18$ & $  2.60 \pm  0.29$ \nl
  988.6 &$ 11.56 \pm  0.46$ & $  4.30 \pm  0.17$ & $  2.78 \pm  0.31$ \nl
  991.7 &$ 11.35 \pm  0.45$ & $  4.15 \pm  0.17$ & $  3.08 \pm  0.34$ \nl
  994.6 &$ 11.37 \pm  0.46$ & $  4.11 \pm  0.16$ & $  2.56 \pm  0.28$ \nl
 1009.6 &$ 10.98 \pm  0.44$ & $  3.76 \pm  0.15$ & $  2.50 \pm  0.28$ \nl
 1012.6 &$ 12.18 \pm  0.49$ & $  3.70 \pm  0.15$ & $  2.98 \pm  0.33$ \nl
 1019.6 &$ 12.75 \pm  0.51$ & $  4.20 \pm  0.17$ & $  3.33 \pm  0.37$ \nl
 1022.6 &$ 12.09 \pm  0.48$ & $  4.57 \pm  0.18$ & $  3.02 \pm  0.33$ \nl
\enddata

\end{deluxetable}

\begin{deluxetable}{lcccccc}
\tablewidth{0pt}
\tablecaption{Variability Parameters}
\tablehead{
\colhead{Subset} &
\colhead{No.\ Epochs}  &
\colhead{$\Delta t_{\rm ave}$ (days)} &
\colhead{$\Delta t_{\rm med}$ (days)} &
\colhead{Mean flux\tablenotemark{a}} &
\colhead{$F_{\rm var}$} &
\colhead{$R_{\rm max}$} \nl
\colhead{(1)} & 
\colhead{(2)} & 
\colhead{(3)} &
\colhead{(4)} &
\colhead{(5)} &
\colhead{(6)} &
\colhead{(7)} 
} 
\startdata
\multicolumn{7}{l}{X-Ray (2--10\,kev):} \nl
All 	& 140 & 7.5 & 7.3 & $6.20\pm3.85$ & 0.618 & $42.85\pm16.36$ \nl
Year 1 	&  33 & 2.8 & 0.7 & $7.87\pm3.25$ & 0.411 & $5.766\pm0.459$ \nl
Year 2  &  20 & 11.5& 13.0 & $6.42\pm4.95$ & 0.770 & $17.68\pm2.03$ \nl
Year 3  &  19 & 14.1 & 14.1 & $1.97\pm1.81$ & 0.906 & $14.79\pm5.65$ \nl
\nl
\multicolumn{7}{l}{Continuum (5100\,\AA):} \nl
All 	& 126 & 7.4 & 3.1 & $12.74\pm1.08$ & 0.077 & $1.397\pm0.079$ \nl
Year 1 	&  51 & 3.9 & 3.0 & $13.66\pm0.82$ & 0.050 & $1.287\pm0.073$ \nl
Year 2  &  38 & 5.8 & 3.2 & $12.16\pm0.76$ & 0.050 & $1.305\pm0.074$ \nl
Year 3  &  37 & 6.9 & 4.5 & $12.06\pm0.69$ & 0.043 & $1.281\pm0.072$ \nl
Subset 1&  29 & 2.8 & 2.2 & $13.38\pm0.92$ & 0.059 & $1.287\pm0.073$ \nl
\nl
\multicolumn{7}{l}{H$\beta$:} \nl
All 	& 126 & 7.4 & 3.1 & $ 4.77\pm0.48$ & 0.095 & $1.592\pm0.092$ \nl
Year 1 	&  51 & 3.9 & 3.0 & $ 4.91\pm0.48$ & 0.092 & $1.423\pm0.064$ \nl
Year 2  &  38 & 5.8 & 3.2 & $ 4.65\pm0.44$ & 0.086 & $1.505\pm0.086$ \nl
Year 3  &  37 & 6.9 & 4.5 & $ 4.50\pm0.42$ & 0.086 & $1.462\pm0.084$ \nl
Subset 1&  29 & 2.8 & 2.2 & $ 5.02\pm0.51$ & 0.096 & $1.389\pm0.062$ \nl
\nl
\multicolumn{7}{l}{He\,{\sc ii}\,$\lambda4686$:} \nl
All 	&  93 &10.1 & 4.5 & $ 3.50\pm0.67$ & 0.154 & $2.038\pm0.316$ \nl
Year 1 	&  29 & 6.9 & 4.0 & $ 4.08\pm0.52$ & 0.064 & $1.660\pm0.258$ \nl
Year 2  &  33 & 6.7 & 3.6 & $ 3.42\pm0.62$ & 0.143 & $1.987\pm0.309$ \nl
Year 3  &  31 & 8.2 & 5.0 & $ 3.04\pm0.38$ & 0.057 & $1.680\pm0.263$ \nl
Subset 1&  17 & 4.8 & 3.5 & $ 4.08\pm0.65$ & 0.114 & $1.660\pm0.258$ \nl
\enddata
\tablenotetext{a}{X-ray fluxes in counts per second in 2--10\,keV band.
Other fluxes in units as in Table 2.}
\end{deluxetable}

\begin{deluxetable}{lcc}
\tablewidth{0pt}
\tablecaption{Cross-Correlation Results and Virial Mass
Estimates}
\tablehead{
\colhead{Parameter}  &
\colhead{$F({\rm H}\beta)$} &
\colhead{$F({\mbox{He\,{\sc ii}\,$\lambda4686$}})$} \nl
\colhead{(1)} & 
\colhead{(2)} & 
\colhead{(3)} 
} 
\startdata
$\tau_{\rm cent}$ (days) & $5.92^{+3.13}_{-1.96}$ & $4.49^{+4.91}_{-5.60}$ \nl
$\tau_{\rm peak}$ (days) & $4.6^{+4.5}_{-1.5}$ & $5.9^{+8.9}_{-4.8}$ \nl
$r_{\rm max}$    & 0.800 & 0.767   \nl
$V_{\mbox{\scriptsize FWHM}}$ (km\,s$^{-1}$) & $1110\pm190$ 
& $5430\pm510$\nl
$M_{\rm vir}$ ($10^6\,M_{\odot}$) & $1.1^{+0.8}_{-0.5}$ &
$19.4^{+21.5}_{-24.4}$ \nl
\enddata
\end{deluxetable}
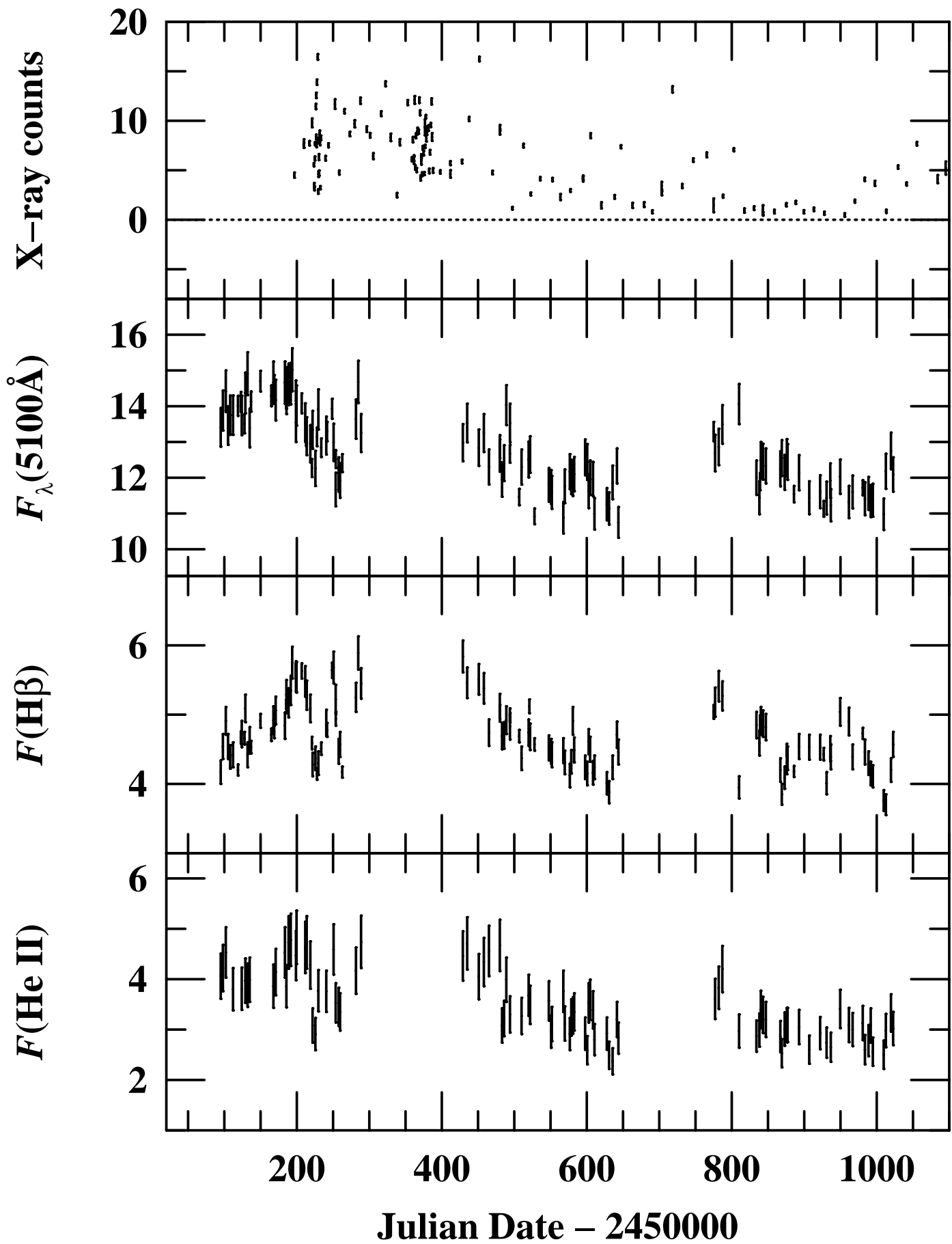
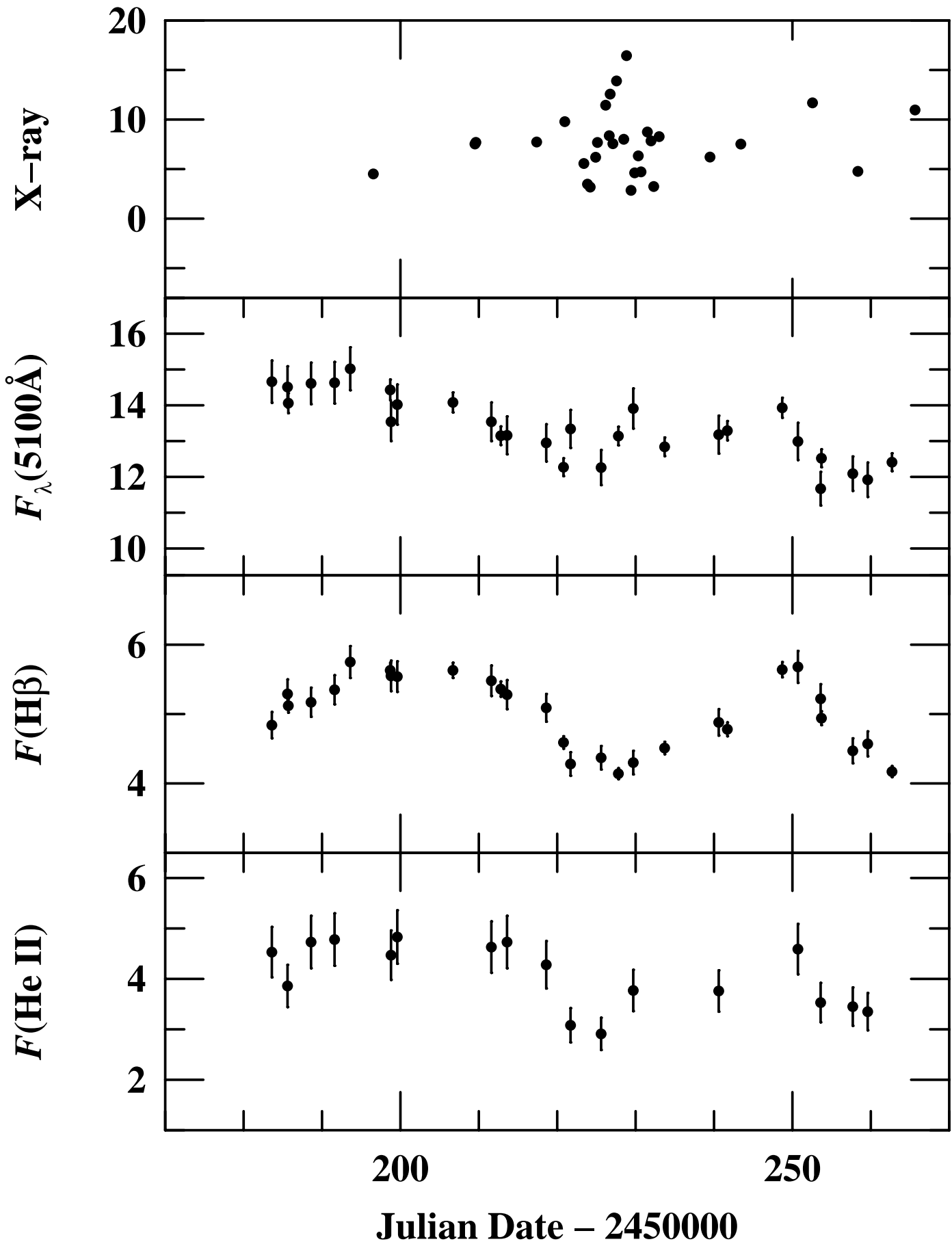
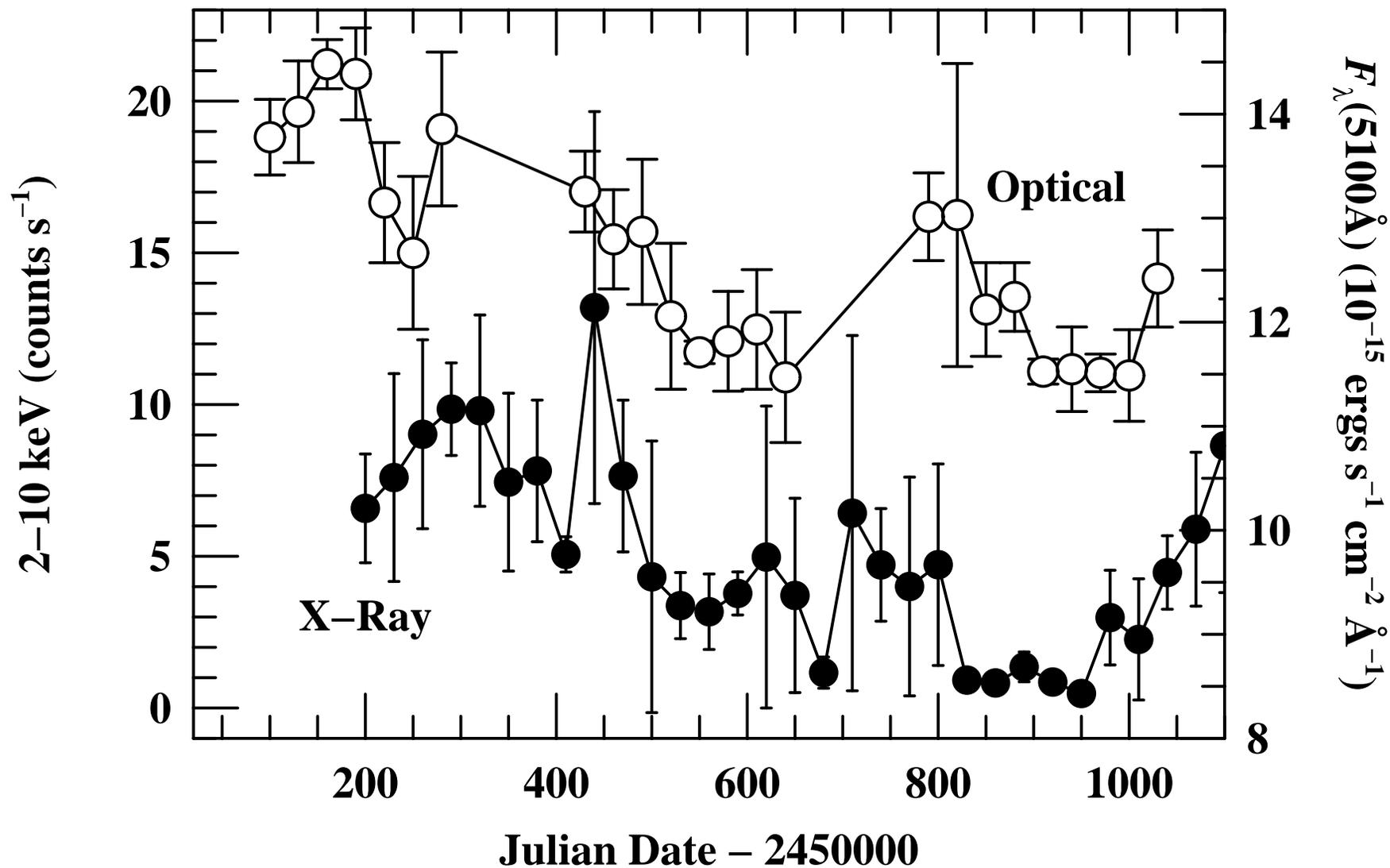
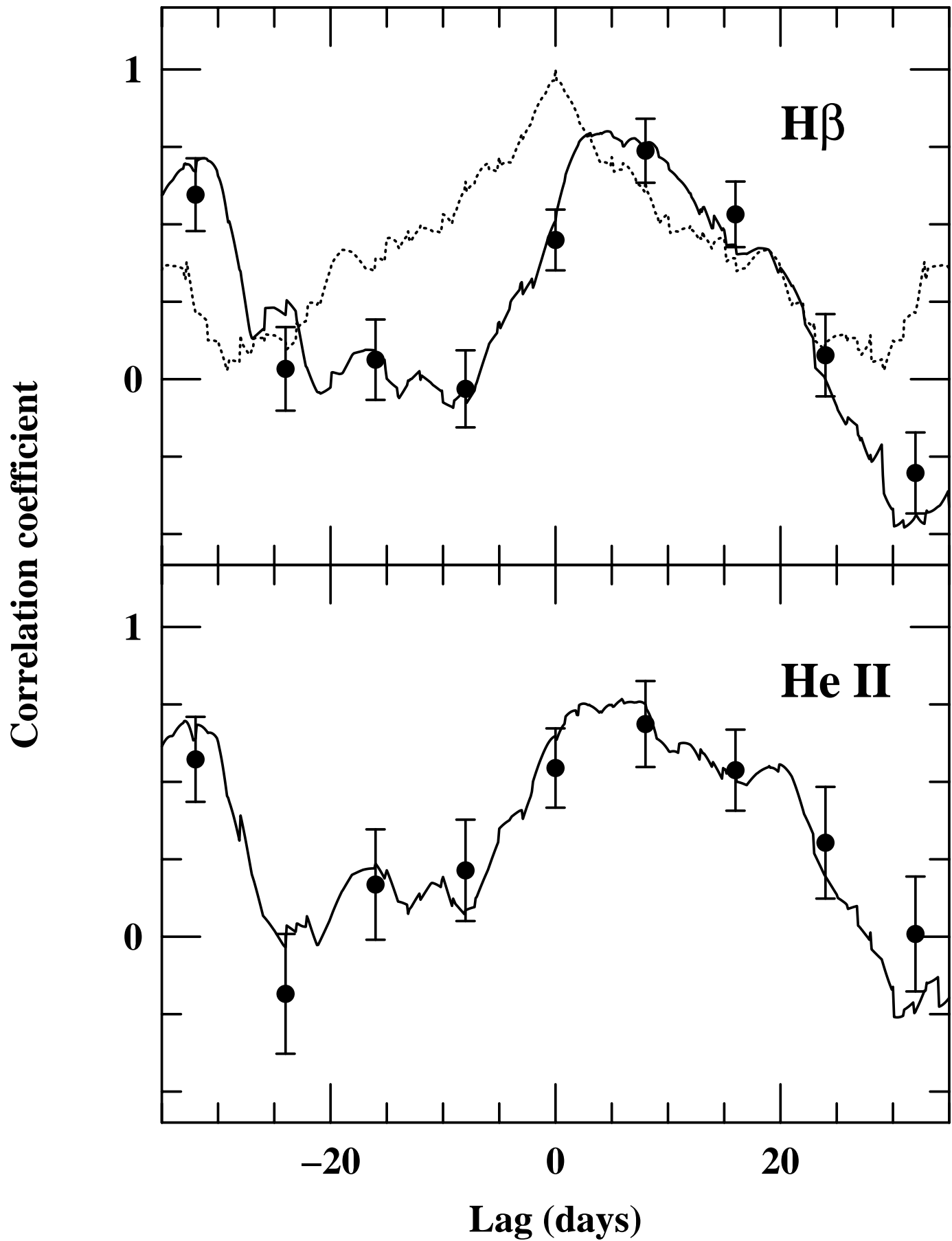
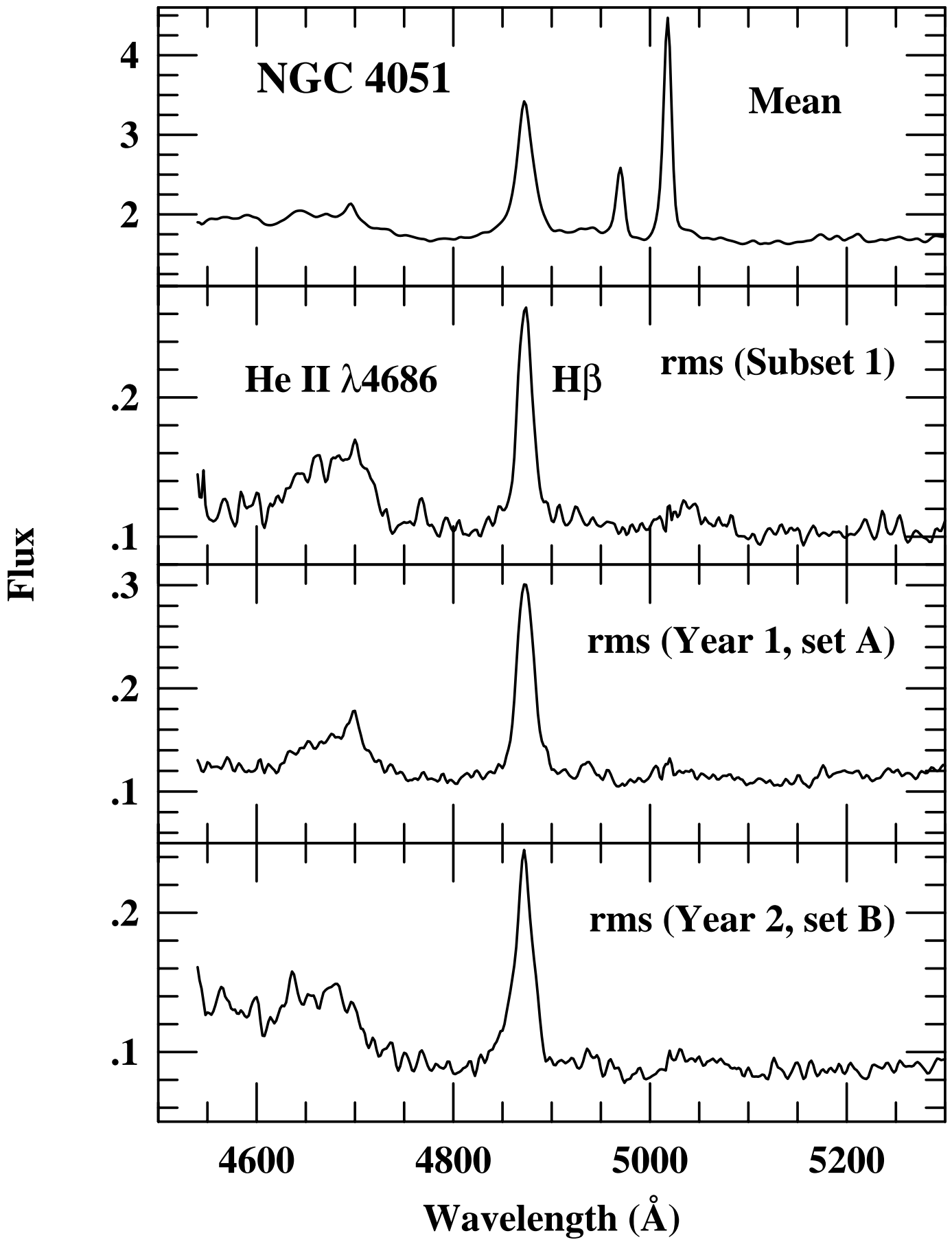
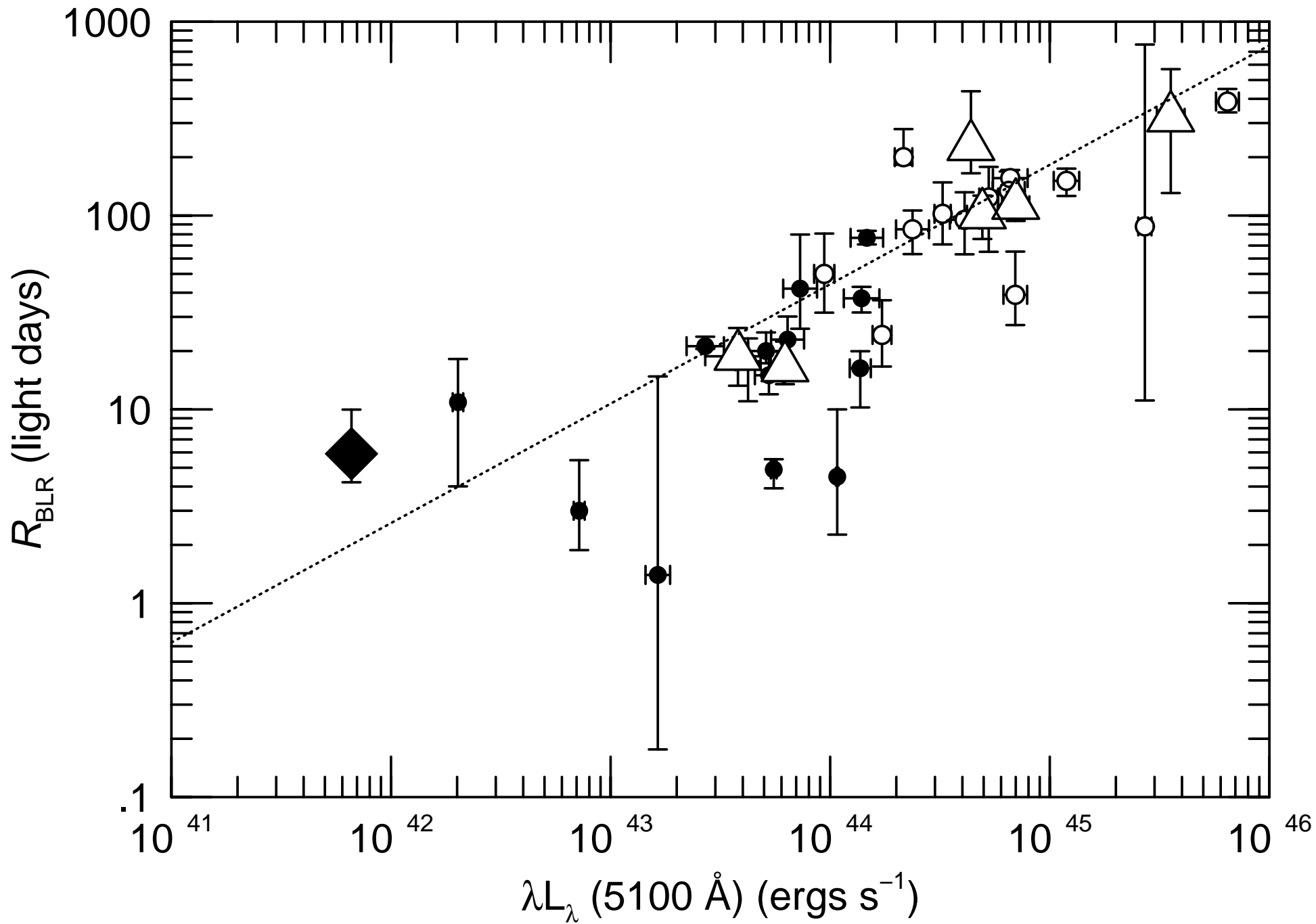
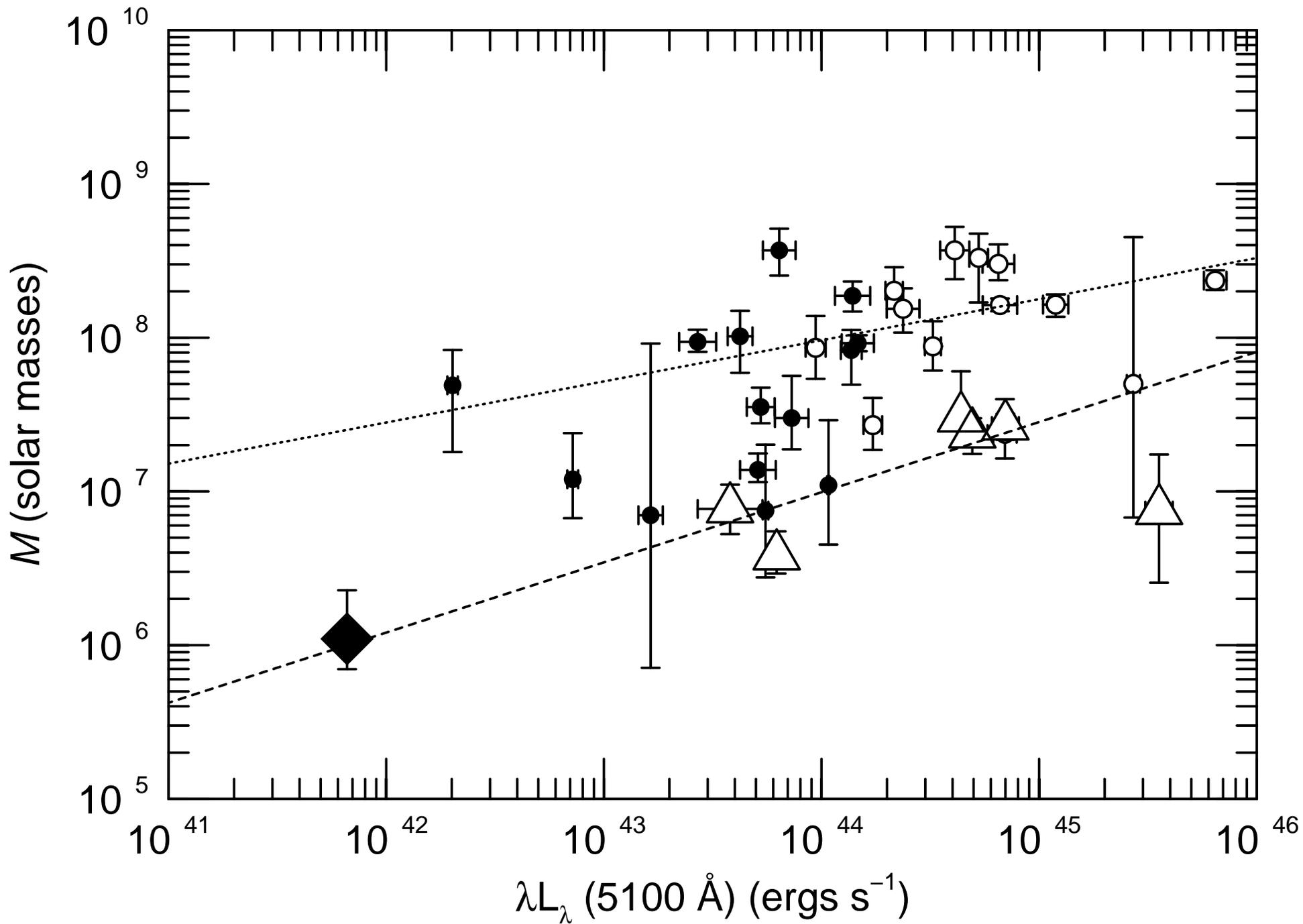
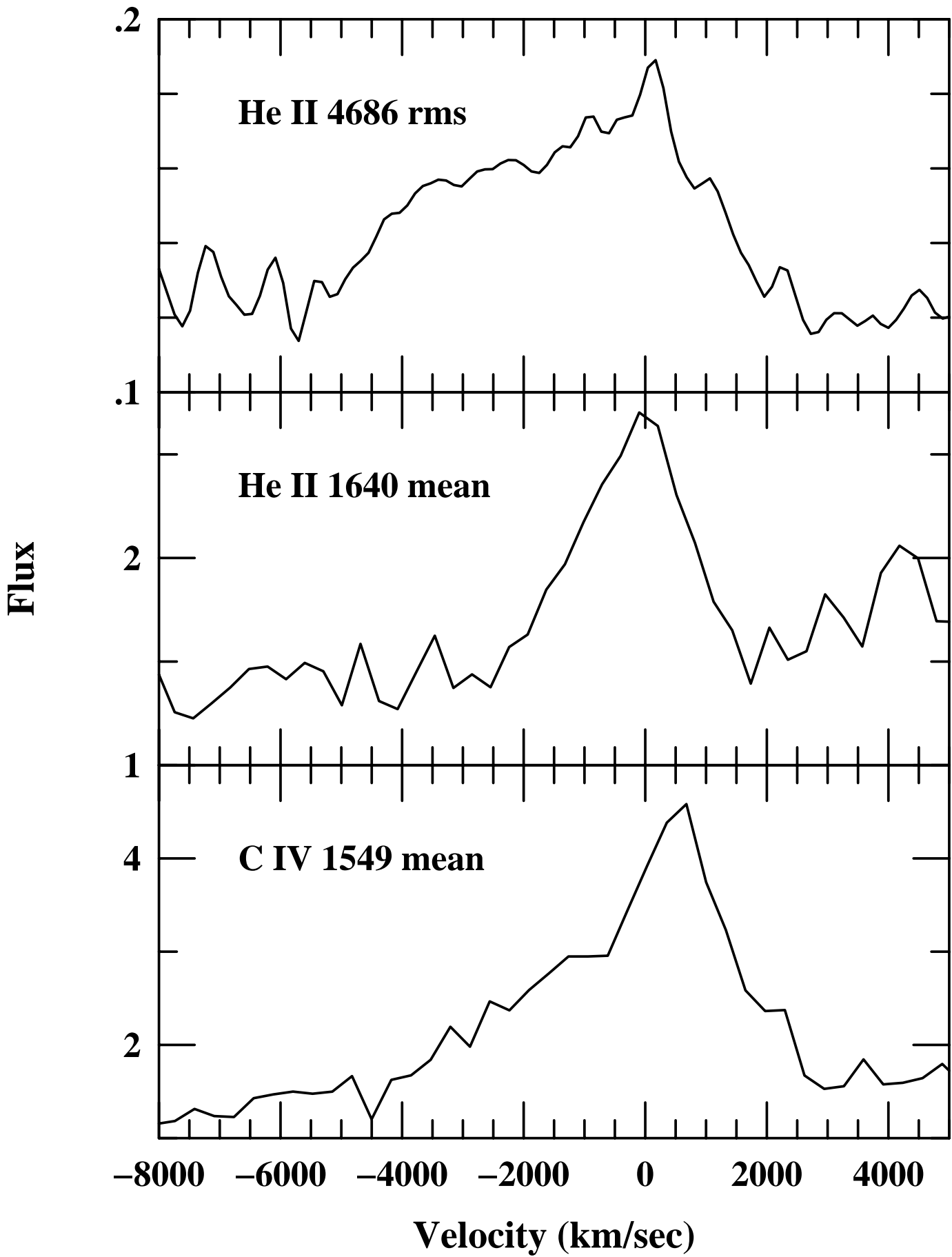
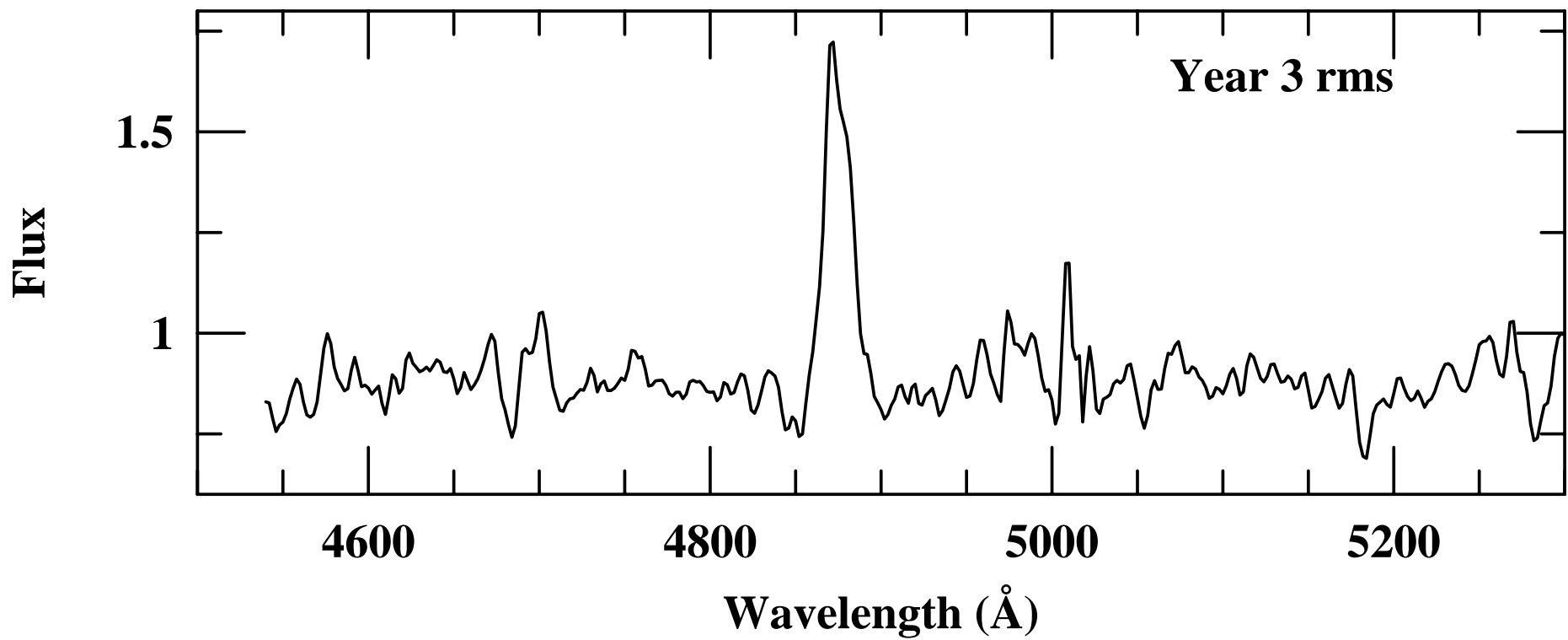
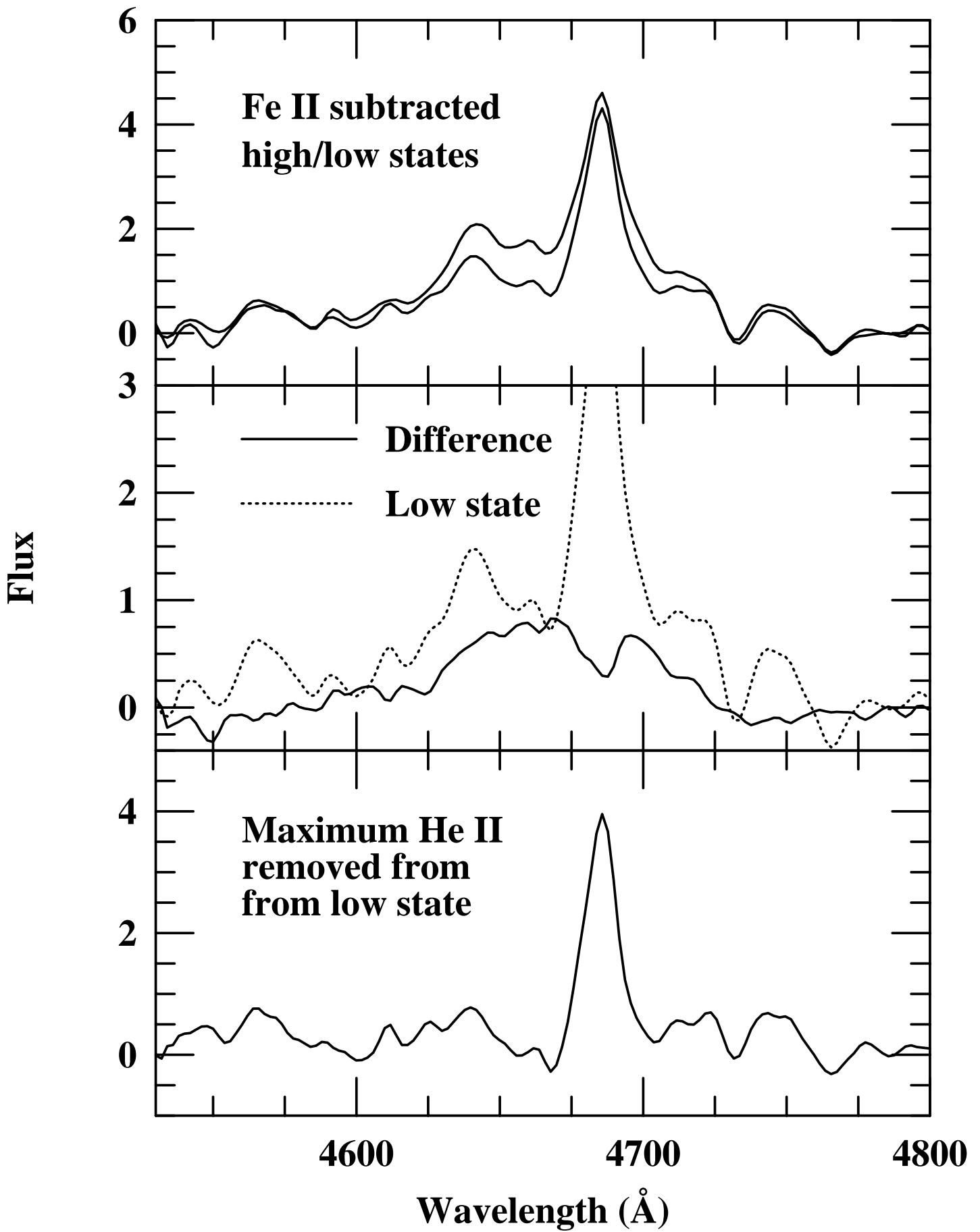
\end{document}